\documentclass[10pt,journal,cspaper,compsoc]{IEEEtran}
\usepackage{url}
\usepackage{subfig}
\usepackage{tikz}
\usepackage{tkz-graph}
\usepackage{wrapfig}
\usepackage{tabularx}
\usepackage{rotating}
\usepackage{amsmath}
\usepackage{amsthm}

\usepackage{setspace}

\begin{document}

\title{Citations, Sequence Alignments, Contagion, and Semantics: On Acyclic Structures and their Randomness}

\author{Sandeep Gupta\thanks{A part of the work was carried out when the author was at San Diego Supercomputer Center, University of California, San Diego, CA}\\
Network Dynamics and Simulation Science Laboratory\\
Virginia Bioinformatics Institute\\
Virginia Tech, Blacksburg, VA 24061\\
sandeep@vbi.vt.edu\\
}

\IEEEcompsoctitleabstractindextext{%
\begin{abstract}

Datasets from several domains, such as life-sciences, semantic web, machine learning, natural language processing, etc. 
are naturally structured as acyclic graphs. 
These datasets, particularly those in bio-informatics and computational epidemiology, have grown
tremendously over the last decade or so. Increasingly, as a consequence, there is a need to  build and evaluate various strategies  for processing 
acyclic structured graphs. 
Most of the proposed research models the real world acyclic structures
as random graphs, i.e., they are generated by randomly selecting a subset of edges from all possible edges. 
Unfortunately the graphs thus generated have predictable and degenerate structures, i.e., the
resulting graphs will always have almost the same degree distribution and very short paths. 

Specifically, we show 
that if $O(n \log n \log n)$ edges are added to a binary tree of $n$ nodes then 
with probability more than $O(1/(\log n)^{1/n})$ the depth of all but $O({\log \log n} ^{\log \log n})$ vertices of the dag collapses to 1.
Experiments show that irregularity, as measured by distribution of length of random walks from root to leaves, is also predictable and small.
The degree distribution and random walk length properties of real world graphs from these domains  are significantly different from random graphs
of similar vertex and edge size.
\end{abstract}
\begin{keywords}
Acyclic Graphs, Graph Generation, Random Graphs,  Semantic Web, Life Sciences, Benchmarking
\end{keywords}}

\maketitle
\IEEEdisplaynotcompsoctitleabstractindextext
\IEEEpeerreviewmaketitle

\section{Introduction}
\label{sec:intro}

Data generators play an important role in algorithm design and optimization engineering. 
They are an important tool for modeling, benchmarking, scalability analysis, and cost estimation. 
In the last decade with the unprecedented growth in Internet, WWW, and social networks the 
need for generators that produce graphs reflecting structures of these domains became  prominent and have been an active area of research.


It was discovered in ~\cite{Faloutsos:1999:PRI:316188.316229,Kumar:2000:SMW:795666.796570} that 
such networks are fractal in nature (scale free) and 
can be described via physical phenomena of ``rich gets richer'' or ``preferential attachment''~\cite{barabasi-1999-286}. The mathematical
model that produces such fractal graphs is based on  R-MAT~\cite{DBLP:conf/sdm/ChakrabartiZF04} or Kronecker Product~\cite{Leskovec:2010:KGA:1756006.1756039} and is the central theory underlying 
the scale free generators.
A large body of work exists that utilizes either of the mathematical models to develop scalable scale free 
graphs~\cite{DBLP:conf/sdm/ChakrabartiZF04,Tangmunarunkit:2002:NTG:964725.633040}. 
These works have contributed significantly towards the development of network protocols, algorithms, and
architecture design. 

The community has paid little attention to generation of acyclic graphs. Acyclic graphs, much like scale free graphs, appear in many areas of computation and engineering. 
Knowledge representation, binary decision diagram, dependency graphs, semantic web, and binaries of computer programs are a few examples.
In the field of life-sciences and bio-informatics, such structures are used to create ontologies that represent the compendium of factual information.
Advancement in  these fields has lead to  an almost 
rapid increase in the number of ontologies.
In the field of 
genome sequencing, the problem of multiple sequence alignment  can be represented as directed acyclic graphs~\cite{lee2002multiple}. 
In computational epidemiology, the graph representing the (plausible) spread of contagion from person to person 
in a population is acyclic~\cite{Marathe:2013:CE:2483852.2483871}. 
Citation network among publications and patents is also acyclic. 

Unlike social networks and WWW, the workloads in life-sciences and knowledge engineering disciplines are much more complex and include 
reachability and pattern queries, and lowest common ancestors~\cite{Dehainsala:2007:OOD:1783823.1783879}. 
It is important for the database and the computing world to be able to develop algorithms over such workloads to better address the needs of the domain science. 
Graphs generators that produce realistic data sets would be indispensable for this exercise. 

\newcommand{\TBD}[1]{\textcolor{red}{#1}}

The size of acyclic graphs used in real world is not yet as large as the social-network graphs (upwards of multi billion edges) but they 
can potentially be many orders of magnitude larger than those present currently. 
This is particularly true in 
life-sciences, bio-informatics and knowledge representation domain. This is because disparate ontologies can reference 
entities across each other. For example, the Gene Ontology, has been constructed by combining ontologies of sub-species. 
Another ontology, Unified Medical Language System (UMLS) which maps the terminology of 60 different biomedical source vocabularies currently consists 
of one million biomedical concepts and five million concept name~\cite{DBLP:journals/bioinformatics/KohlerPL03}.

\begin{figure}
  \includegraphics[width=3.00in]{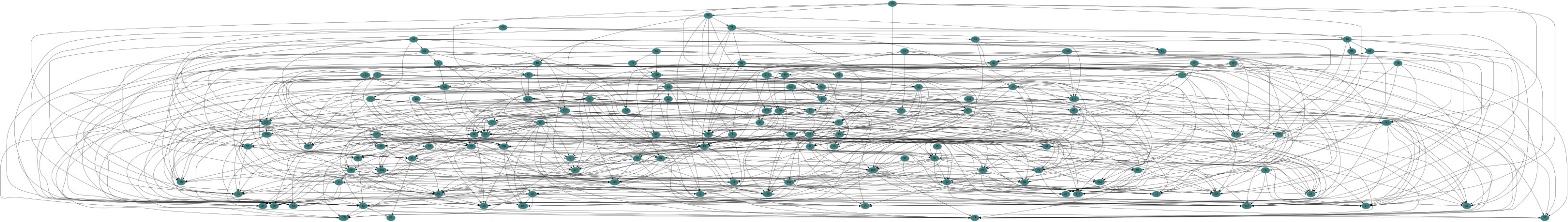}
  \caption{{\small Drawing of 128 node and 611 edges graph generated by the random generator using the Graphviz graph layout toolkit~\cite{EllsonGKNW00}. The aspect ratio of the figure is determined 
by graphviz based up on the structural properties of the graph. The small aspect ratio of the figure clearly indicates that the graph has collapsed; more so 
when compared with the non-degenerate (and asthetically sound) height to width ratio of the drawing of the graph generated using our method (see figure~\ref{fig:s7_graphviz}). \label{fig:rdag_s7_graphviz}}}
\end{figure}

\begin{figure}
 \includegraphics[width=3.00in]{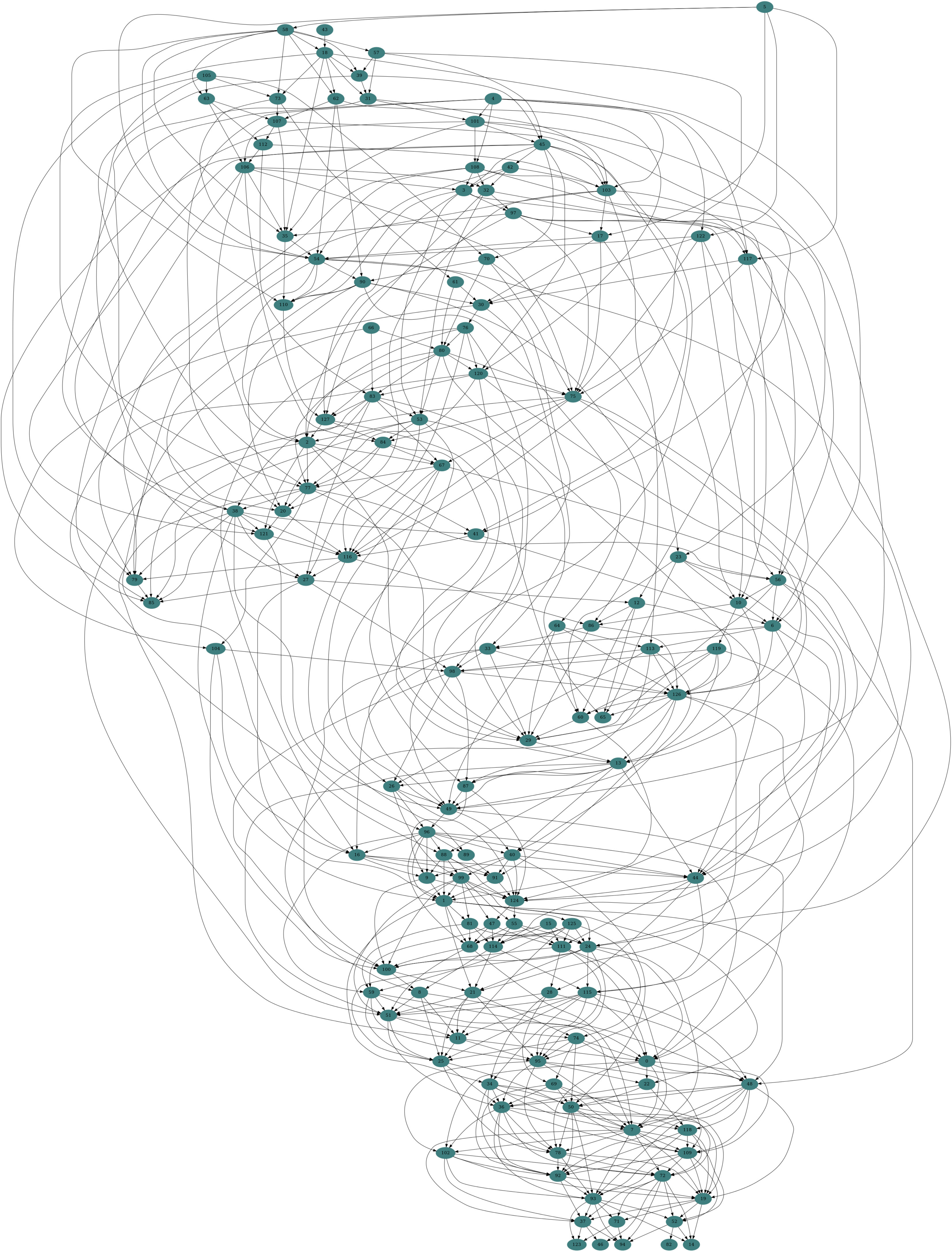}
 \caption{{\small Graphviz drawing of 128 node and 556 edges non-``random'' graph. The height to width ratio of the figure is almost
one which is a further evidence that the graph has non-degenerate structure. \label{fig:s7_graphviz}}}
\end{figure}

Most of the generators currently used to build acyclic graphs do so via random selection of edges~\cite{jin2008efficiently,wang2006dual}
Unfortunately, the graphs {\it thus generated  almost always collapses}. Figure~\ref{fig:rdag_s7_graphviz} is an illustration 
of a randomly generated graph using the popular graph drawing tool, Graphviz~\cite{EllsonGKNW00}. Figure~\ref{fig:s7_graphviz} is provided
for comparison; it  illustrates a ``non''-radom acyclic graph with same number of nodes and edges. The aspect ratio and other visual attributes
of the figures are automatically
determined by the Graphviz. We can see that the random graph in figure~\ref{fig:rdag_s7_graphviz}  has a collapsed (or close
to collapsed) structure. On the contrary, the graph in figure~\ref{fig:s7_graphviz} is asthetically more sound non-degenerate structure.

In this work, we first give the
intuitive meaning of collapse by drawing analogies from other domain of sciences and then provide
two formal metric to measure collapsness on acyclic graphs.

Being in (or reaching) a collapsed state implies having (or arrived at) a degenerate structure, i.e., lacking
any order or property. In other words, irrespective of the generation process
the resulting structure  ''looks'' and behaves the same. This phenomean of leading to collapsed structure has counterpart in
 science and engineering. In mathematics, singularity theory is a special discipline which studies ``failure of manifold structure''. 
It is an important tool in the study of general relativity which studies how strong gravitational fields may change the structure of space time.

One of ways  singularity may arise is due to degeneration. Degeneration is a state in which a class of objects change their nature 
so as to belong to another, usually simpler class. For example, point is a degenerate case of the circle as radius approaches zero and circle 
is a degenerate for ellipse as ecentricity approaches zero. We demonstrate in this paper that an analogous phenomena
happens to  acyclic graph generation via random selection of edges, i.e., the resulting graph belongs to a simpler 
and trivial subset of a much larger domain of all acyclic graphs.

\subsubsection*{Two measures over acyclic graphs to capture collapsness}
In this paper we formulate two measures using which we  quantify the
degree of ``collapsness'' in the acyclic graphs.
First,  called ``depth distribution'' measure computes 
the distribution of  the shortest distances of the leaves of the dag to the roots. 
The analogous of this measure in general undirected graphs is the measure called ``diameter'', which is
computed as the maximum over the shortest path between all pairs.

This measure is easy to compute and correlates with  ``collapsenss'', i.e.,
depth distribution in a collapsed graph will be restricted to a narrow range of values. 
However, it is not a  true indicator of collapsness, i.e., an acyclic graph with depth distribution within
a narrow range of values is not necessarily a collapsed graph. 

Second measure is termed Random Root-to-Leaf  path length (or RRL) measure. RRL is  a distribution obtained from a random variate. Its value is assigned 
as the length of a random walk from root to the leaf. The distribution over this random variate 
accurately captures ``collapsness''. If significant fraction of the paths (99\% or above) 
have length less than $\beta$, where $\beta$ is a  very small value (typically within 10),  then we say that the acyclic graph
is ``collapsed''.

\subsubsection*{Large scale naturally occurring acyclic graphs}
We require large acyclic graphs extracted from real data in order to demonstrate that they are non-degenerate and cannot be modeled using  random graph generator. 
This task turns out to be  challenging inspite of large acyclic graphs naturally occurring in many domains. It turns out that for most domains, inferring  the acyclic structure is  either prohibitively expensive or not feasible at all due to lack of tools and technology. 

For example, large acyclic graphs are embedded in contagion processes that model the spread of disease in a population~\cite{barrett2007modeling}. Such processes consists of a collection of agents. An agent is in healthy, infected, or, cured state. The process starts with a small subset of population being in infected state. Infection gets transmitted from infected populace 
to healthy population as and when they come in close proximity of each other. 
As the time progresses, the new infected agents together with the  previously infected 
agents infect remaining population.
This process repeats until the contagion
reaches a pandemic state or is curtailed by external interventions. Since a agent can only get infected once, because once cured he becomes immune to the disease, 
the infectee (agent T) - infected (agent I) relationship forms an acyclic graph. The infectee - infected edge, $T \rightarrow I$,  captures 
fact that there is a likelihood that $I$'s infection may be have been passed from agent $T$. 
In other words, in the resulting acyclic graph, parents of 
agent $I$ represent all agents through which I could have been infected. 
Characteristic of this contagion graph for a particular disease (such as influenza) would be of great use to epidemiologist and public policy experts in order to understand the evolution of the disease in 
a population.

However, building such a graph would require keeping track of every individual and the state of their health which is not feasible due to   logistical constraints. 

Similarly, the graph representing the evolutionary relationship among different species is an acyclic graph. This is because in addition to orthologs (genetic parent of a species) and paralogs (genetic siblibs of the species),  species 
such as, Corynebacterium diphtheriae ,  have rearticulation and horizontal gene transfer belonging to different parent species. 
However, most biology research ignore  evolutionary relationships due to horizontal gene transfer and have represented  them  as (phylogenetic) trees. Again, obtaining genetic markers of species and building all evolutionary relationships including mining for horizontal gene transfer is a undertaking beyond the scope of this work.

Ontologies also happen to encode acyclic graphs. Ayclic structures via is-a and part-of relationships over the domain vocabulary form the backbone of the ontology over which rest of the ontological constructs and relationships are defined.  The Unified Medical Language System (UMLS) is a ontology which maps the terminology of 60 different biomedical source vocabularies currently consists 
of one million biomedical concepts and five million concept name~\cite{DBLP:journals/bioinformatics/KohlerPL03}. Unfortunately, they haven't yet connected the vocabularies via is-a, part-of, and other ontological relationships.

We have been able to obtain large acyclic graphs from three domains: string matching data structure called 
deterministic acyclic finite state automaton (DAFSA), the open-cyc ontology, and the partonony relationships over dictionary words in wordnet.  We shall study various graph characteristics, in particular ``collapseness'', of these graphs  and compare them with the characteristics obtained from random graphs.

\begin{figure}[!ht]
\centering
\includegraphics[height = 1.0in,width=1.0in,angle=-90]{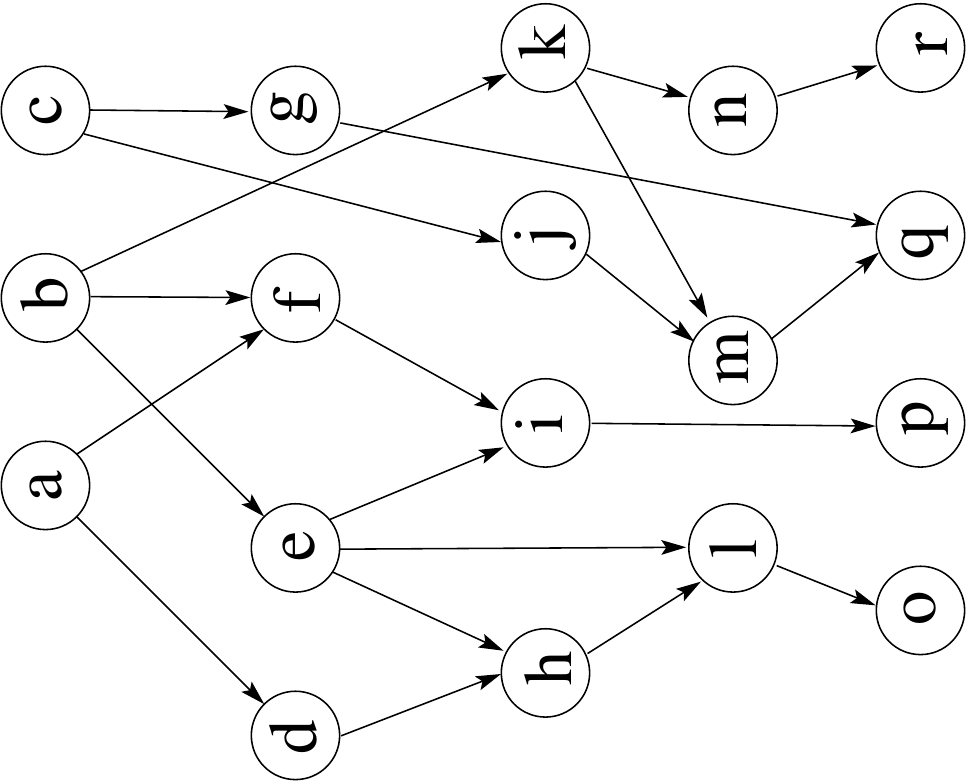}
\caption{A sample DAG}
\label{fig:exDAG}
\end{figure}

\section{Existing Graph Generators}
The simplest graph generator is due to Erdos-Renyi~\cite{Erdos:1960} which generates a random graph. Let $n,p$  be the
number of nodes and the edge probability, respectively. The generator by Erdos-Renyi creates a graph with $n$ nodes and (expected) $np$ number of edges.
An edge is created by  picking two nodes at random and joining them. They show that even such a simple generator exhibits an interesting phenomena of phase change. Namely, there exists a narrow range for $p$ values for which the number of connected components drop significantly for very small increment in $p$. 
Planar triangulated graphs~\cite{springerlink:10.1007/978-3-642-21204-8_39} are another class of random graph generators in which the points are embedded on a Euclidean plane. The Delaunay triangulation of the points yields a random graph. 
The R-MAT or Kronecker product based graph generators, mentioned in the introduction, is by far the most widely used graph generator. It has been extensively used to generate scale free graphs.
All of these generators are meant for creating random or scale free, directed or undirected graphs.

In~\cite{DBLP:conf/odbis/TheoharisGC07}, the authors presented a linear programming based approach for 
generating acyclic graphs. In their approach, each node is a variable in the linear programming formulation.
While the graphs generated using this approach are rich, a significant limitation of this algorithm 
is that it does not scale to multi-millions node acyclic graphs due to the extensively large size of the 
linear programming systems involved. 
The other difficulty with this approach is 
that it consists of many constraints. Setting up the constraints  such that a solution to the linear program exists is non-trivial.
Another approach for generating acyclic graphs was presented in~\cite{Melancon:2000:RGD:868981}, where the authors
strove to generate acyclic graphs with given number of nodes uniformly at random. Their goal was to develop 
test suites for graph drawing packages and focused mostly on graphs of sizes in the range of 100--1000 nodes.
The limited dataset size feasible through this procedure hints that purely random generation of acyclic graph
is computationally very expensive. 

In addition to work on generating  graph structures, there has been a surge of interest for benchmarks in social media analytics, semantics web, and bio-informatics, all of
which deal with graph structured datasets. As a consequence, modeling and generation of graph recently has been active topic of research~\cite{zhang2012srbench,bail2010justbench,duan2011apples,bock2008benchmarking}. 
As noted earlier, graphs in social media can be modeled as R-MAT or Kronecker graphs. Various work address generation of such graphs (see ~\cite{Lothian_Powers_Sullivan_Baker_Schrock_Poole_2013} and references within). 
Contrary to social media domain, which has focused primarily on structural property of the graph, the field of semantic web has addressed
the problem of benchmarking and graph generation in wide variety; work in~\cite{morsey2011dbpedia} propose real world datasets (along with collection of sparql queries)
as a benchmark. In~\cite{guo2005lubm} authors develop a benchmark that includes a data generator and a set of queries.
The data generator not only include structural attributes of the graph but also generates labels for the nodes and edges. The objective is
to create dataset that closely models real world semantic web dataset. Finally, works in~\cite{schmidt2009sp,bizer2009berlin} address the choice of sparql queries for benchmarks.

The most popular scheme for generating acyclic graphs is to start with a simple acyclic structure (e.g. chains or binary tree)  and add edges between nodes 
$(u,v)$ if the level of $u$ is less then the level of $v$. The nodes $u, v$ in the node pair are selected randomly. As mentioned earlier in section~\ref{sec:intro},
graphs generated using such techniques, though scalable, have an inherent tendency to collapse.
In the rest of this section, we pause briefly introduce notations (to be used throughout the paper) and then proceed to discuss the limitations of random acyclic generator. 

\subsection{Notations, Terminologies, and Definitions}
\label{subsec:notations}
Given a directed graph $G=(V(G), E(G))$, let $n = \#V(G)$ be the number of nodes and $E(G)$ denote the number of edges in $G.$ 

A path is a sequence of edges, 
\begin{equation}
\vec P = \langle (v_1= u , v_2), (v_2,v_3) \dots, (v_{t -2}, v_{ t -1}), (v_{t-1}, v_t = v) \rangle.
\nonumber
\end{equation}
Alternatively, the same path can be represented as a sequence
of vertices $\vec P = \langle v_1,  v_2, \dots, v_{{t}} \rangle.$
We say that a path is simple if a node appears at most once in the sequence; else
the path has cycles. Graph $G$ is a  {\bf directed acyclic graph} (DAG) if all paths
are simple. Hence, nodes in acyclic graphs naturally induce a topological ordering. Let $T(u)$ denote
the  rank of node $u$ induced by the topological ordering. 

An edge $e \in E(G)$, in acyclic graphs, 
is said to be out-incident on node $u$ and in-incident on node $v$ if $e = (u,v).$
Further, $u$ is termed as an {\bf in-incident} neighbor (or parent)
of $v$ and $v$ is termed as an {\bf out-incident} neighbor (or children) of $u.$ The total number of in-incident neighbors of a node $u$ is termed as it's in-degree, denoted 
by $I(u).$ Similarly, the total number of out-incident neighbors of $u$ is termed as it's {\bf out-degree}, denoted by $O(u).$  
Nodes with in-degree zero are termed as root nodes while nodes with out-degree zero are leaf nodes. 
We call the rest of the nodes as DAG  nodes. The rest of the notations will be defined as and when required during algorithm description.
Figure~\ref{fig:exDAG} shows an example DAG with 
$parents(e) = \{b\}$ and $children(e) = \{h,l,i\}.$


\subsection{Collapsing Nature of Acyclic Graphs}
\label{sec:collapse}
With very high probability random acyclic generators yield bipartite graphs irrespective of the initial 
structure and the type of random number generator. We demonstrate this phenomena when the initial structure is a  binary trees. 
Let $B_T$ be a binary tree with $n$ nodes and $n-1$ edges. Let $td_v$ denote the tree depth of node $v$ in this tree.
We follow the convention that root is at level $0$. The level is in ascending order along the child/descendant axis.
To this tree we add $cn(1+ \epsilon)$ edges for some fixed $c$ and $0 < \epsilon < 1,$ resulting in average degree of the generated graph to be $c(1+\epsilon).$ 
 Each edge $e=(u,v)$ is directed from the source $u$ to $v$ if $td_u < td_v.$
Let $dd_v$ denote the depth of the node $v$ in the graph {\it after} addition of random edges, $dd_v$ being defined as 
\begin{equation}dd_v = min\lbrace dd_u+1 |  u \text{ is-a parent of v}\rbrace.
\label{eq:def_depth} 
\end{equation}

\newtheorem{lemma}{Lemma}[section]

\begin{lemma}
\label{lem:lem1}
Let $(u,v)$ be randomly selected pair under the constraint that $td_u < td_v$. Let $E_{v}$ denote the event that  $td_u < td_v-1.$
Then $Pr(E_v) = \frac{1}{2}$
\end{lemma}
\begin{proof}
Let $k$ be the number of nodes in the tree at depth~$=td_v - 1.$ Then, the total number of nodes with depth~$< td_v - 1$ is $k-1$ 
(by property of binary trees). This implies that $P(td_u < td_v - 1) = P(td_u = td_v-1) = \frac{k}{2k-1} = 1/2,$ for all practical purposes.
Hence, 
$P(E_{v}) = \frac{1}{2}.$
\end{proof}

Suppose the generation is performed in $c$ iterations and in each iteration $n(1 + \epsilon)$ edges are added.
The source and destination nodes are selected randomly.
We say that a node is {\em selected} in a iteration if it is a target of any of the $n(1+\epsilon)$ randomly selected edges. 
\newtheorem{claim}{Claim}[section]

\begin{claim}
In a iteration, with high probability (w.h.p) almost every node selected as a target of a randomly selected edge.
\end{claim}
\begin{proof}
 Let $\bar X_{\gamma}$ denote that $\gamma$ nodes are not selected as target during an iteration. We are required to show that $Pr[\bar X_{\gamma}]$ approaches $0$ from some small value of $\gamma$.

Consider a node $u.$ Let $X_u^i$ denote  the event that the node is selected as target node for the  $i^{th}$ randomly generated edge. 
Let $X_u = \Sigma_{i\in [1:(1+\epsilon)n]} X_u^i$ denote the event the $u$ is selected {\em at least} once as destination and $\bar{X_u}$ denote
the event that $u$ is never selected target i.e $Pr[\bar X_u] = 1 - Pr[{X_u}]$.

Since there are $n$ possible choices for choosing a target node it follows that $Pr[X_u^i] = 1/n$ and $Pr[\bar {X_u^i}] = 1 - 1/n.$
Hence, 
\begin{equation}
Pr[\bar{X_u}] = \Pi_{i \in [1:n(1+\epsilon)]} Pr[\bar{X_u}] = (1-1/n)^{n(1+\epsilon)}
\end{equation} 

The probability that a given set of $\gamma$ nodes, $u_1, u_2, \dots, u_{\gamma}$ are never chosen as target is 
\begin{equation}
\Pi_{j \in [1:\gamma]} Pr[\bar{X_{u_j}}] = (1-1/n)^{n\gamma(1+\epsilon)}
\end{equation}

Since there are ${{n}\choose{\gamma}}$ ways of selecting $\gamma$ points, the probability that $\gamma$ nodes
remain unselected comes out to be:
\begin{equation}
{{n}\choose{\gamma}} (1-1/n)^{n\gamma(1+\epsilon)}
\end{equation}

Simplifying the equation using elementary estimates we have

\begin{equation}
\begin{array}{r c l}
Pr[\bar X_{\gamma}] & = & {{n}\choose{\gamma}} (1-1/n)^{n\gamma(1+\epsilon)} \\
 & <&  (\frac {en} {\gamma})^{\gamma} e^{-{\frac {1} {n}} n\gamma (1 + \epsilon)} \\
&<& (\frac {en} {\gamma})^{\gamma} \frac {1} {e^{\gamma (1 + \epsilon)}}\\
&<& \frac {n^{\gamma}} {{\gamma}^{\gamma}} \frac {1} {e^{\gamma \epsilon}}\\
\end{array}
\end{equation}
Assuming $\gamma = \epsilon = \log n$ we get 
\begin{equation}
\begin{array}{r c l}
{{n}\choose{\gamma}} (1-1/n)^{n\gamma(1+\epsilon)} 
 & <&  \frac {n^{\log n}} {{\log n}^{\log n}} \frac {1} {e^{{\log n}{\log n}}}\\
&<& \frac {n^{\log n}} {{\log n}^{\log n}} \frac {1} {n^{{\log n}}}\\
&<& \frac {n^{\log n}} {{\log n}^{\log n}} \frac {1} {n^{{\log n}}}\\
&<& \frac {1} {{\log n}^{\log n}}\\
\end{array}
\end{equation}
\end{proof}
The above derivation states that if in each round $n \log n $ edges are added 
then the probability that more than $\log n$ nodes are not assigned new edges is less than $O(\frac {1} {{\log n}^{\log n}})$. Since this expression quickly approaches  zero we conclude that w. h.p. almost all nodes will be the target for at least one of the newly inserted edge.

An alternative bound for this can be obtained by applying Chernoff bound~\cite{chernoff1952measure} on $\bar X = \Sigma_{u \in V} \bar X_u $. 
Since $Pr[\bar X_u] = (1-1/n)^{n(1+\epsilon)}$ it follows that the expected value $\mu$ of $\bar X$ is $n * Pr[\bar X_u] = n(1-1/n)^{n(1 + \epsilon)}$. By, Chernoff inequality 

$$Pr[\bar X > \mu (1 +  \delta)]  < ( \frac {e^{\delta}} { {(1 + \delta})^{1 + \delta}})^{\mu}$$
Let $(1+\delta) = \beta = \log \log n$ and let $ \epsilon = \log n - 1$.
Again, using basic estimates we obtain
\begin{equation}
\begin{array}{r c l}
  \mu(1 + \delta) & = & n(1-1/n)^{n(1+\epsilon)} (1+ \delta) \\ 
                   & < &  n e^{-(1+\epsilon)} \beta \\
                  & < &  n e^{-\log n} \beta \\
                  & < & \log \log n \\
\end{array}
\end{equation}

Focusing on the RHS side of the Chernoff expression
\begin{equation}
\begin{array}{r c l}
  \frac {e^{(\beta - 1) \mu}} {\beta^{\beta \mu}} &<& \frac {e^{\beta -1}} {\beta^\beta}\\ 
                             & < &   \frac {e^{\log \log n}} {{e \log \log n}^{\log \log n}} \\
                             & < &   \frac {\log n}  {e{\log \log n}^{\log \log n}} \\
\end{array}
\end{equation}

In other words, if $n \log n$ new edges are added to the graph then the probability that the number of unselected nodes is more than $\log \log n$ is less than $\frac {\log n}  {e{\log \log n}^{\log \log n}}$.

\begin{claim}
  With high probability $dd_v < \max(2, \frac {td_v} {2^c}).$
\end{claim}

\begin{proof}
{\it (Sketch)}
Consider $B_T$ with $n=2.$ This tree has only two nodes, one being the root, the other being the leaf.
Hence, adding a new edge does not induce any reduction in depth. Therefore, 
\begin{equation}
dd_v = td_v = 1 < 2.
\label{eq:twonode}
\end{equation}
Now we consider $B_T$ with $n > 2.$
Through claim 2.1 we know that if in each iteration $n(1 + \epsilon)$ edges are added, where $\epsilon > \log n -1$
then  almost every node 
with high probability (w.h.p) is a target node in at least one of the inserted edges. 
Let $c$ such iterations be performed. 
Let $v$ be such a ``selected'' node. Let $u$ be the corresponding new parent. 
Next, from equation~(\ref{eq:def_depth}), 
$dd_v \le dd_u +1.$
By construction,
\begin{equation}
dd_u \leq td_u.
\label{eq:fact2}
\end{equation}
Using lemma~(\ref{lem:lem1}) and equation (\ref{eq:fact2}), we get,
\begin{equation}
dd_v \leq td_u + 1 < (td_v - 1)+1 = td_v \textrm{ with probability } \frac{1}{2}.
\end{equation}That is,
\begin{equation}
dd_v < td_v \textrm{ with probability } \frac{1}{2}.
\label{eq:eachnode}
\end{equation} 
Equation~(\ref{eq:eachnode}) says that every time an edge is added to a node, the node suffers at least a unit decrease in depth with probability $1/2$.
Since there are $\log n$ edges per node (on average) we can derive a better bound 
on the probability that the node will suffer depth reduction. This is because
for the node to  not suffer depth reduction all the edges should have source node 
from depth $td_v -1$. The probability of this is ${1/2}^{\log n} = 1/n$. Alternatively, we can say that with probability $1-1/n$ the node suffers a depth reduction.

Furthermore since (almost) every edge suffers a depth reduction it follows that the DAG depth of node $v$ would be at least  half of $td_v$ i.e $dd_v < td_v/2$ at the
end of the iteration with probability at least $1-1/n$.

Applying $c$ iterations of edge insertions, will give
\begin{equation}
dd_v < td_v/2^c \quad \textrm{with probability } (1-1/n)^c.
\label{eq:morenodes}
\end{equation}
Equations~(\ref{eq:twonode}) and (\ref{eq:morenodes}) together yield the claim.
\end{proof}
Since the depth of a binary tree is $\log n,$ if we perform merely $\log(\log n)$ iterations, we get
$dd_v < \frac{td_v}{2^{\log\log n}} < \frac{\max{td_v}}{2^{\log\log n}} = \frac{\log n}{\log n} = 1.$
That is, the entire tree (but for $ {\log \log n}^{\log \log n}$ nodes w.h.p) will collapse to a single depth bipartite graph with probability $O{(1-1/n)}^{\log \log n} < O(e^{-\log\log n/n}) < {1/\log n}^{1/n}$

In section~\ref{sec:randomdag} we verify this phenomena for graphs generated using~\cite{acyclic_graph_generator}, which is a random acyclic graph generator 
similar to the one described above, except that they start with random sequences of nodes (instead of a tree) 
and add random edges of form $(u,v)$ if the rank of $u$ is less than the rank of $v$. 

The collapsing phenomena, demonstrated here for binary trees, occurs irrespective of the initial shape of the tree.
Benchmarks based on such generators would simply test set-membership and set intersections and would not stress any graph centric capabilities of the applications, kernels and algorithms.
Many real world acyclic graphs have average degree of more than 1. For e.g., citation datasets such as arxiv (from arxiv.org) and 
 pubmed (from \url{http://www.ncbi.nlm.nih.gov/pmc/});  semantic knowledge database from ~\url{www.mpi-inf.mpg.de/yago-naga/yago/};  gene ontology terms along with their annotation  from ~\url{www.uniport.org} 
have average degrees of 11.12, 4.45, 6.38, and 4.99, respectively~\cite{Yildirim:2010:GSR:1920841.1920879}.


\begin{table*}[ht]
\begin{tabular*}{0.986\textwidth}[c]{| l | l | l | l | l | l |}
\hline
DAG & num nodes & edge factor & (root,leaf) factor & out degree (mean, var)  & in degree (mean, var)\\
\hline
WORDNET & 74374 & 2.04 & (0.77,.00001) & (1.02,0.02) & (1.02,34.55) \\
CYC &  116482 & 7.71 & (.81,0.0005)  & (3.85,8.35) & (3.85,26535) \\
CIT & 205952 & 6.07 & (0.42,0.31) &  (3.04,106.40) & (3.04,22.31) \\
DWAG & 1502025 & 4.78& (.00004,\~0.0) &  (2.39,8.77) & (2.39,141497) \\
PATENTS & 3774768 & 8.75 & (0.14,0.44) &  (4.37,60.58)  & (4.38,44.30) \\
\hline
\end{tabular*}
\caption{Characteristics of acyclic graphs obtained from real world\label{tbl:acyc_real}}
\end{table*}

\begin{table*}[ht]
\begin{tabular*}{1.05\textwidth}[ht!]{| l | l | l | l | l | l |}
\hline
DAG & num nodes & edge factor & (root,leaf) factor & out degree (mean, var)  & in degree (mean, var)\\
\hline
WORDNET-RAND & 55814 & 2.66 & (0.69,0.17) & (1.33, 1.00) & (1.33,5.32) \\
CYC-RAND &  113100 & 6.17 & (.80,0.059)  & (3.09,3.466) & (3.08,42.30) \\
CIT-RAND & 204786 & 6.03 & (0.42,0.41) &  (3.04,5.51) & (3.01,9.52) \\
DWAG-RAND &1459895   & 4.18 & (.09,0.30) &  (2.09, 6.08) & (2.09,1.89) \\
PATENTS-RAND & 3768265 & 8.01 & (0.14,0.55) &  (4.00,13.29)  & (4.00,6.5) \\
\hline
\end{tabular*}
\caption{Characteristics of graph generated using random selection of edges\label{tbl:acyc_rand}}
\end{table*}

\begin{figure*}[ht!]
 \captionsetup[subfigure]{labelformat=empty}
\subfloat[]{%
      \small
      \begin{tabularx}{\textwidth}{p{1cm}XXX}
        & Indegree Distribution & Outdegree Distribution & RRL path length distribution
      \end{tabularx}
    }

 \subfloat[]{
   \includegraphics[width=1.50in,angle=-90]{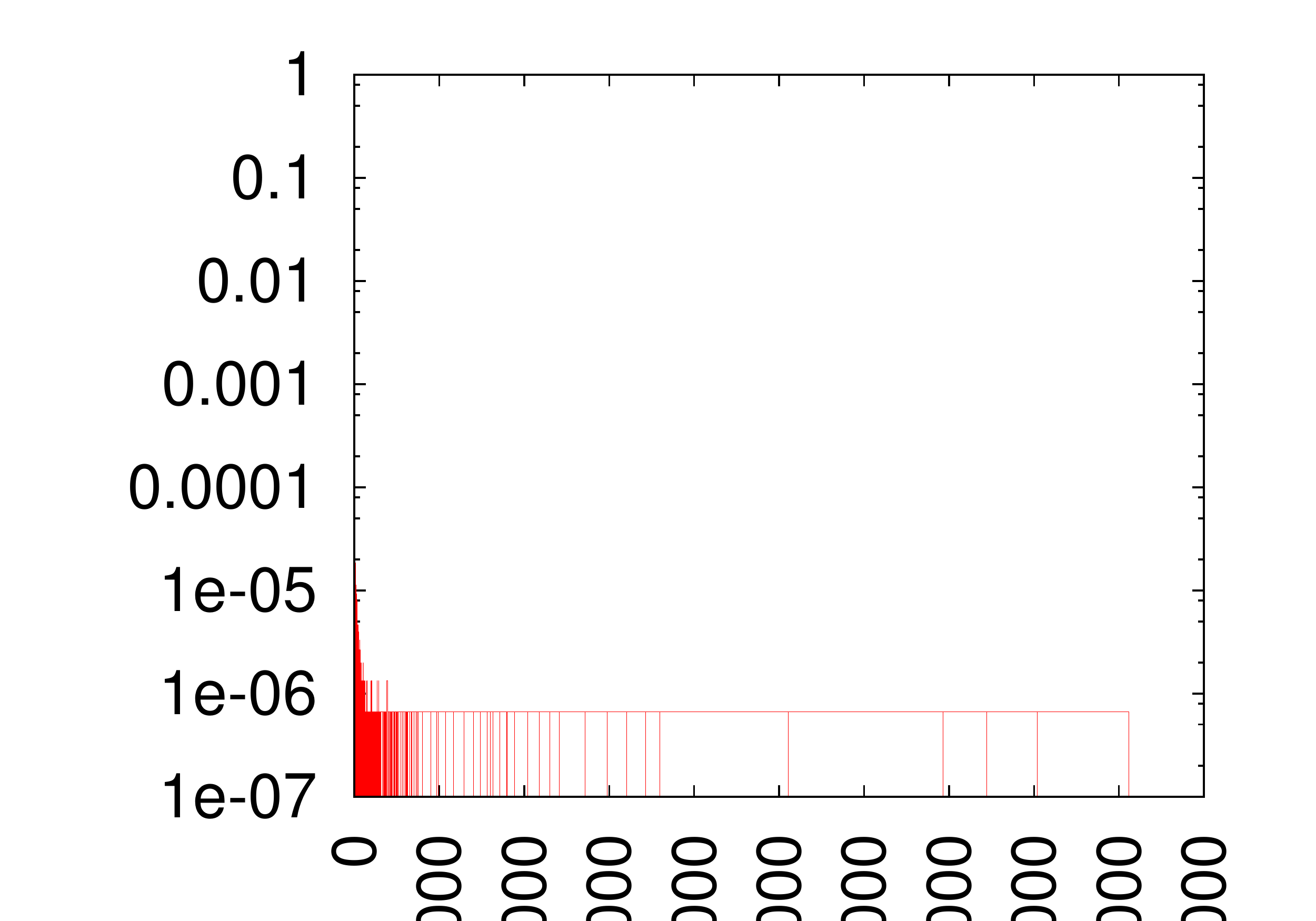}
}\hspace{-1em}
 \subfloat[]{
   \includegraphics[width=1.50in,angle=-90]{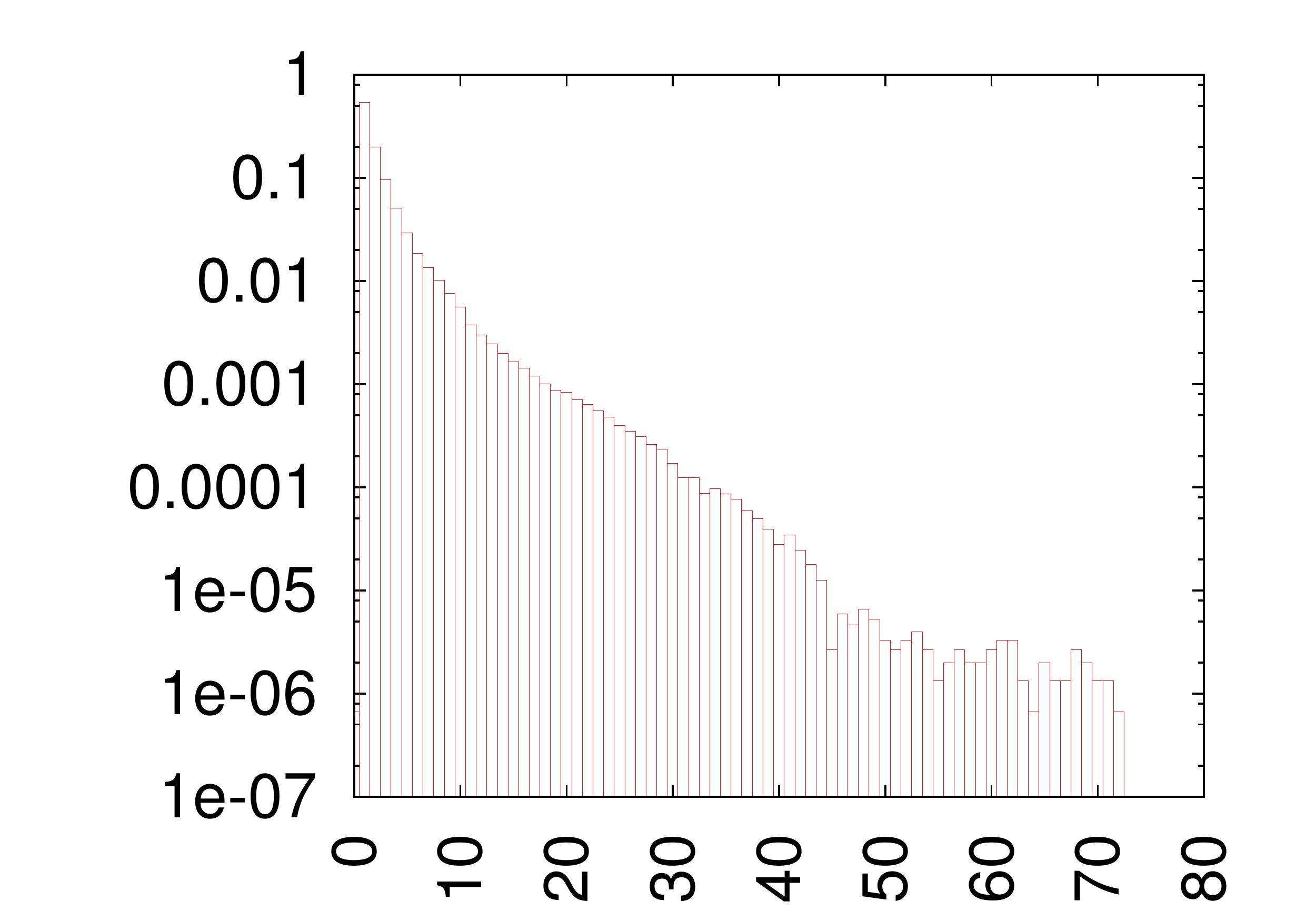}
}\hspace{-2em}
 \subfloat[]{
   \includegraphics[width=1.50in,angle=-90]{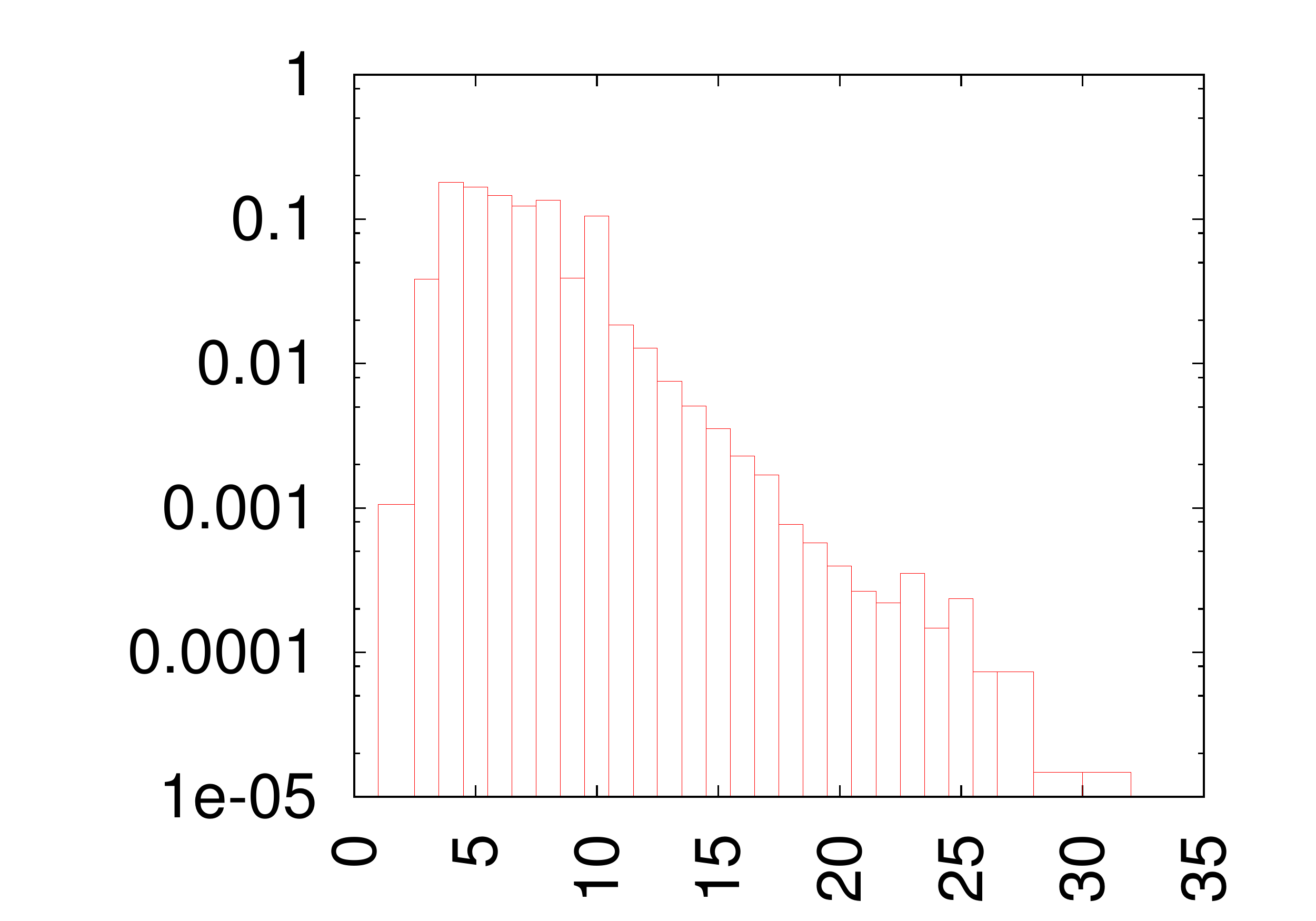}
   {\rotatebox{-90}{\kern1.0cm DWAG}}
}

 \subfloat[]{
   \includegraphics[width=1.50in,angle=-90]{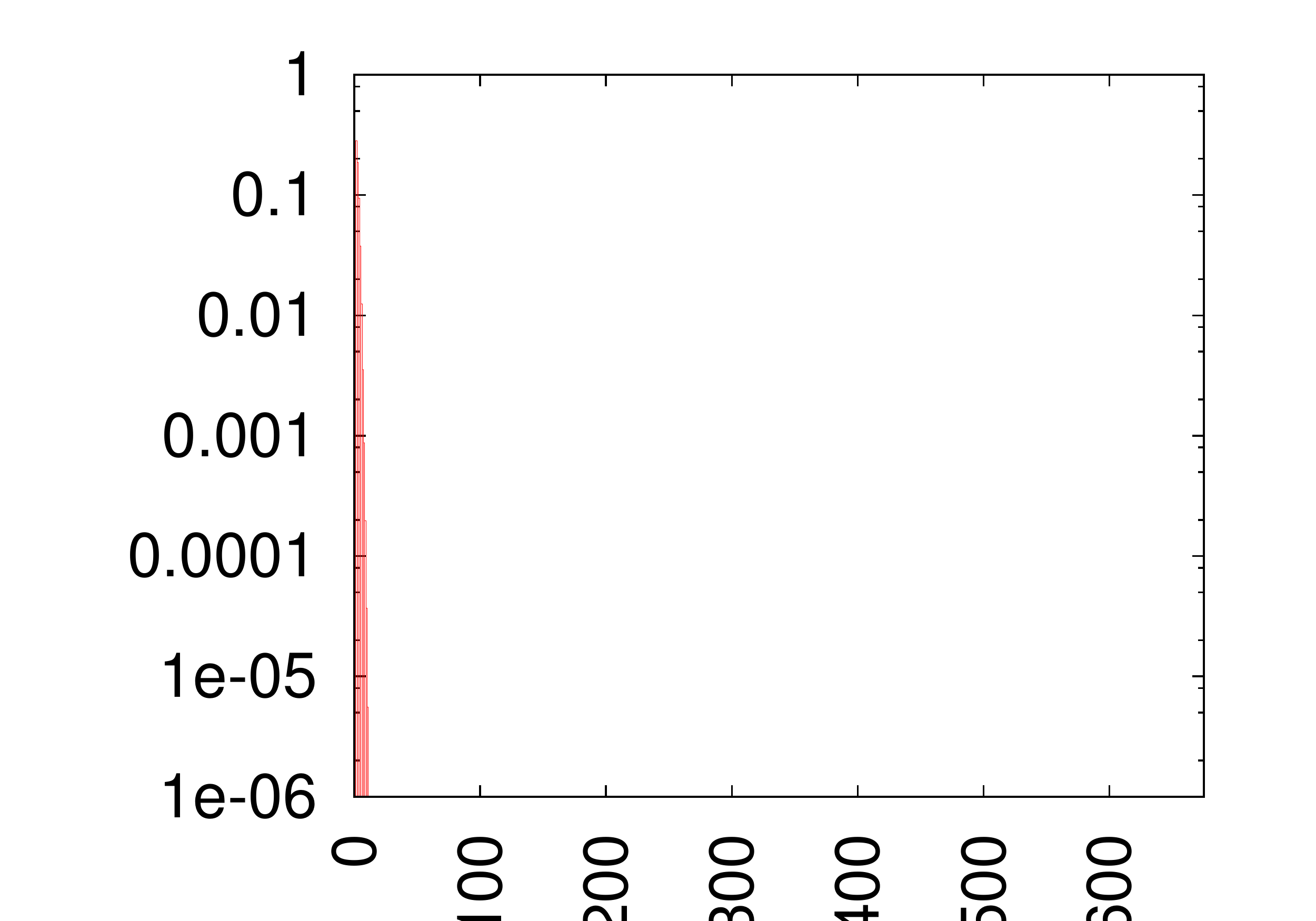}
}\hspace{-1em}
 \subfloat[]{
   \includegraphics[width=1.50in,angle=-90]{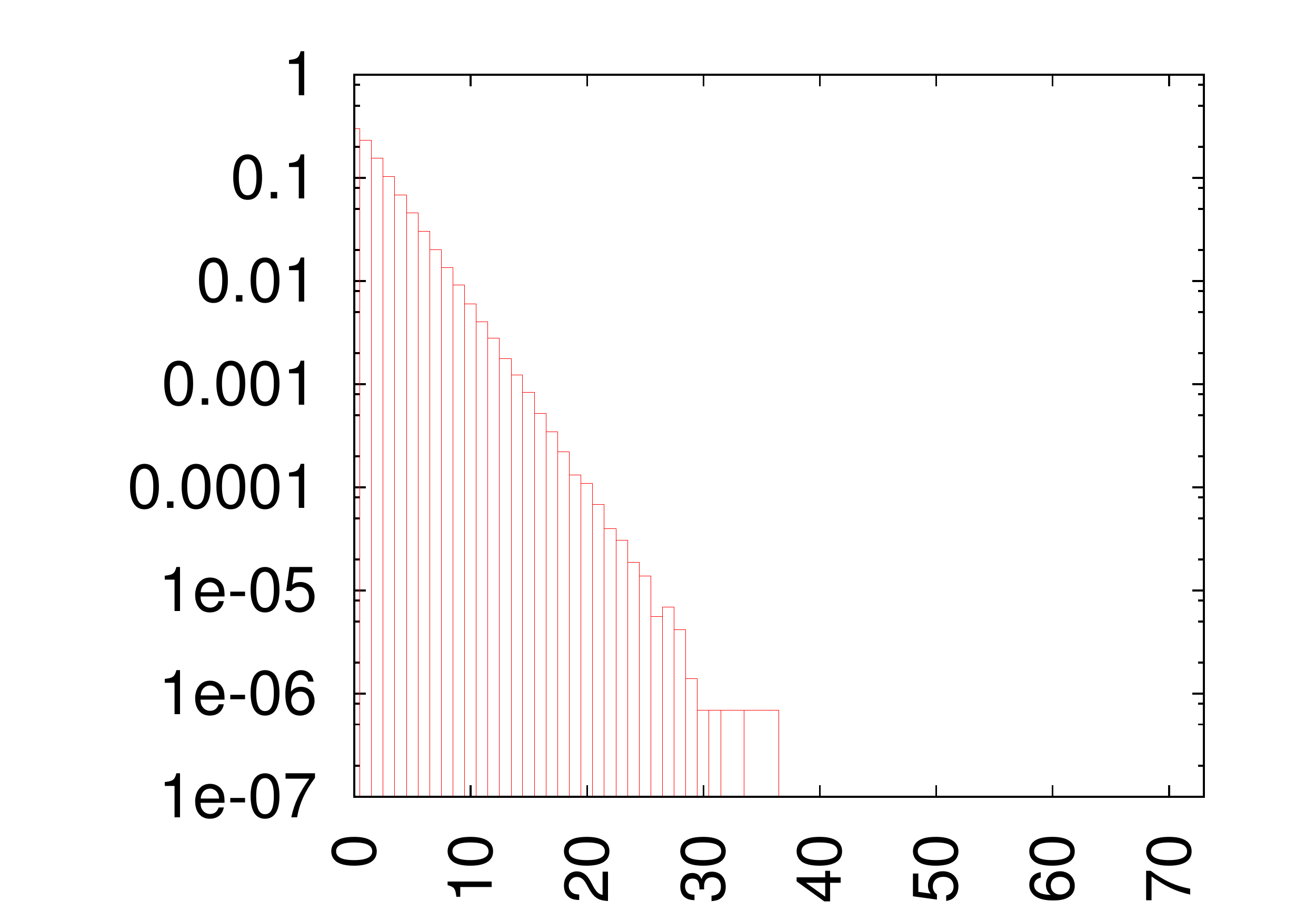}
}\hspace{-1 em}
 \subfloat[]{
   \includegraphics[width=1.50in,angle=-90]{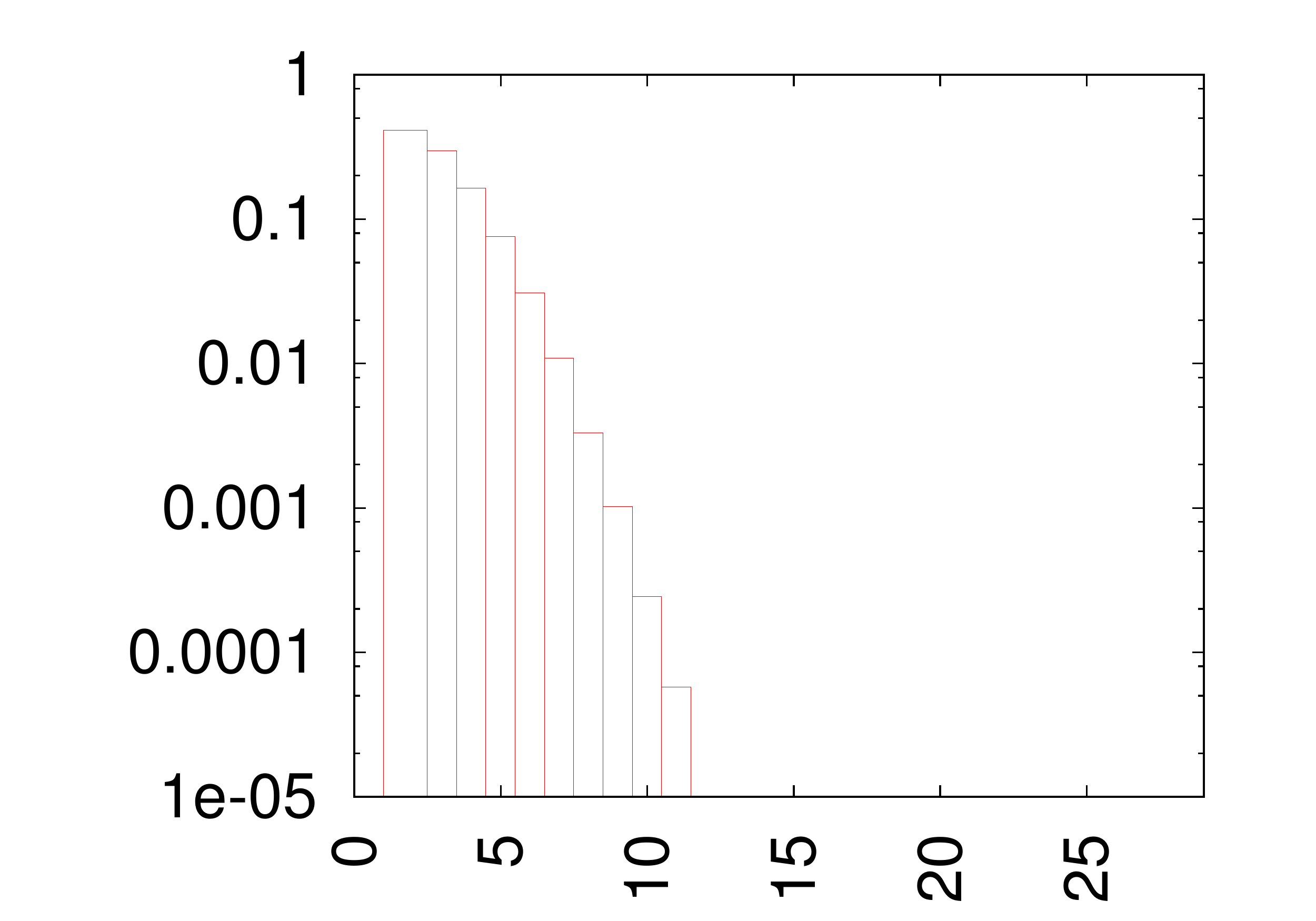}
   {\rotatebox{-90}{\kern0.2cm DWAG-RANDOM}}
}

 \subfloat[]{
   \includegraphics[width=1.50in,angle=-90]{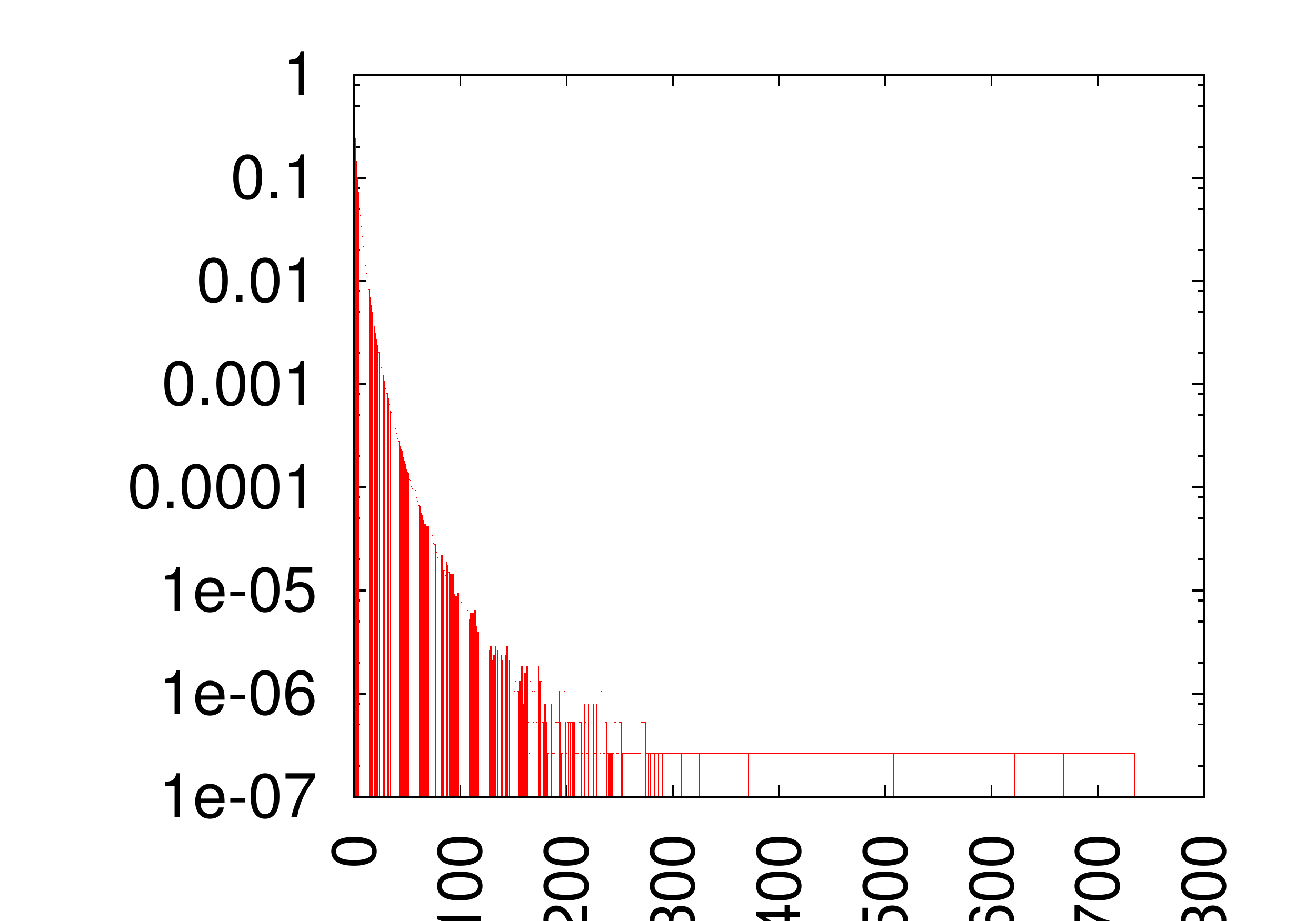}
}\hspace{-1em}
 \subfloat[]{
   \includegraphics[width=1.50in,angle=-90]{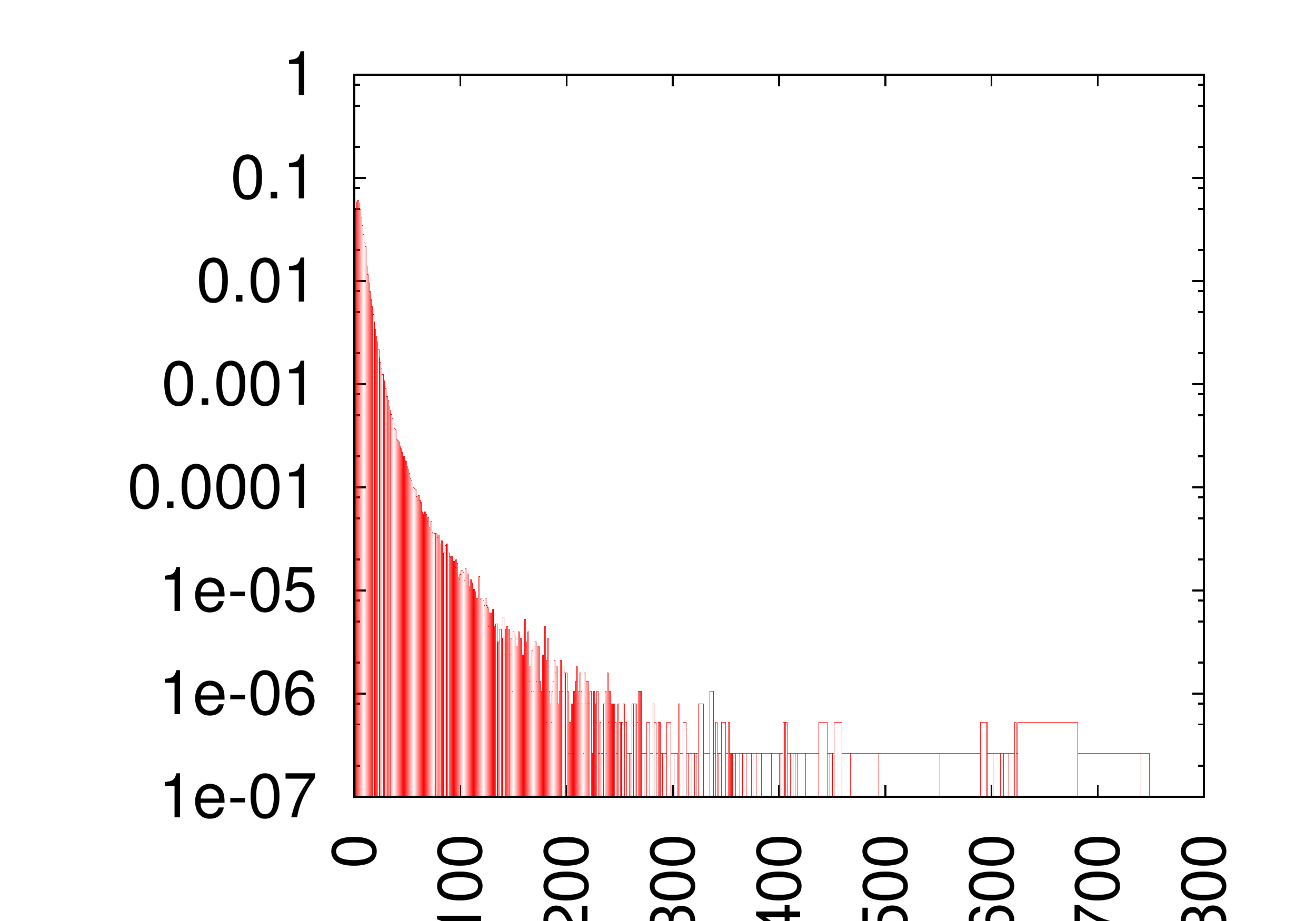}
}\hspace{-1em}
 \subfloat[]{
   \includegraphics[width=1.50in,angle=-90]{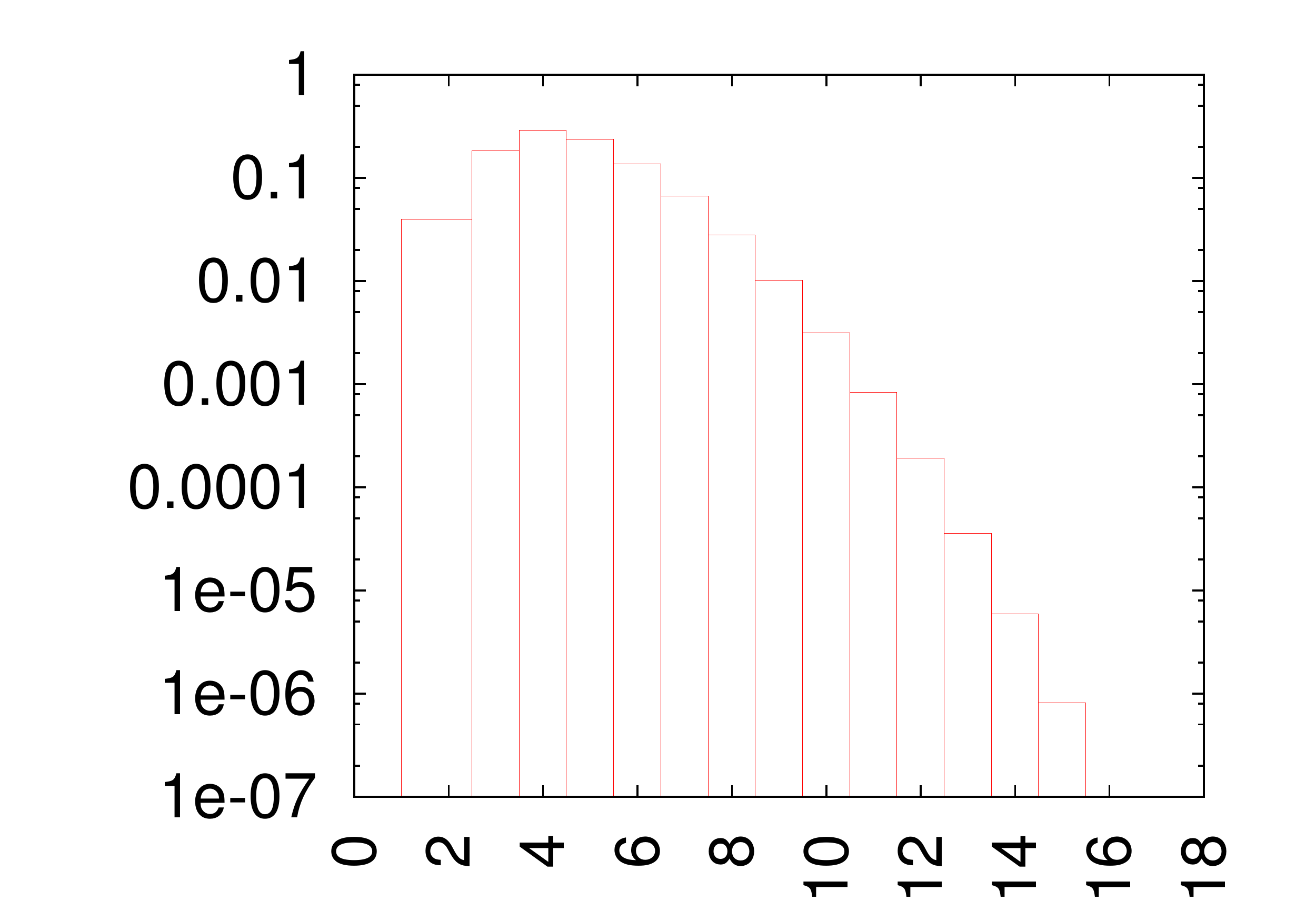}
    {\rotatebox{-90}{\kern0.8cm PATENTS}}
}

 \subfloat[]{
   \includegraphics[width=1.50in,angle=-90]{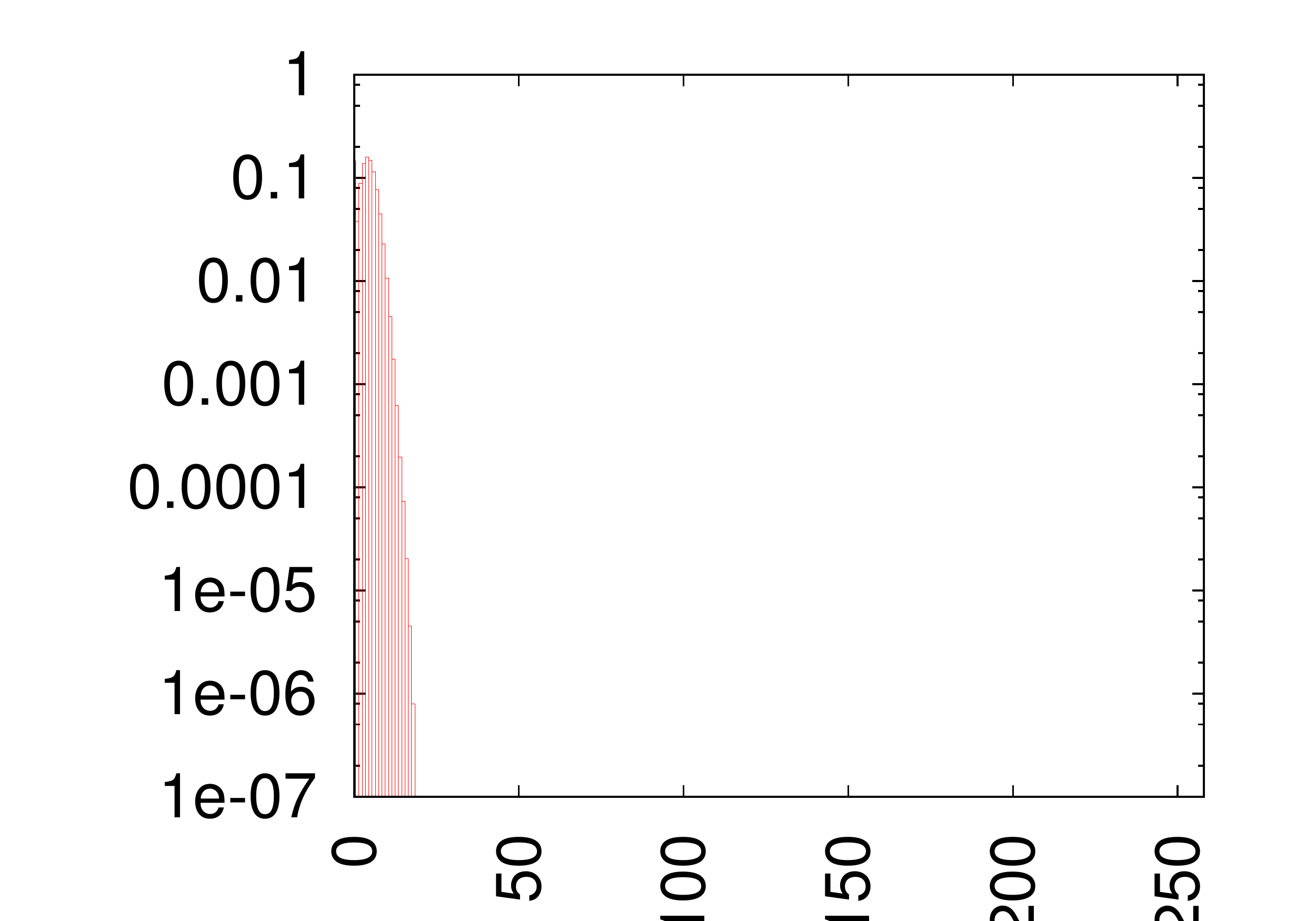}
}\hspace{-1em}
 \subfloat[]{
   \includegraphics[width=1.50in,angle=-90]{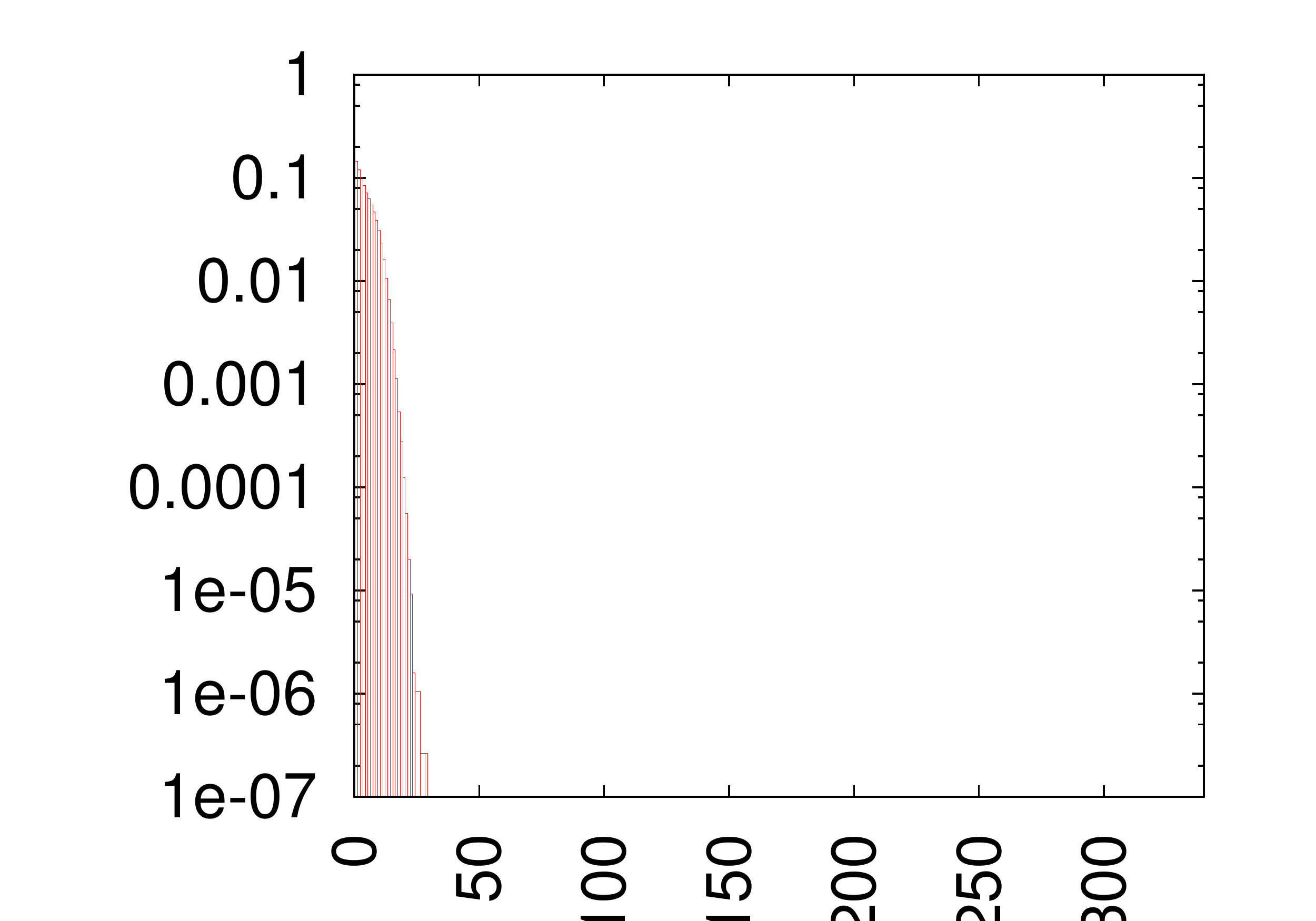}
}\hspace{-1em}
 \subfloat[]{
   \includegraphics[width=1.50in,angle=-90]{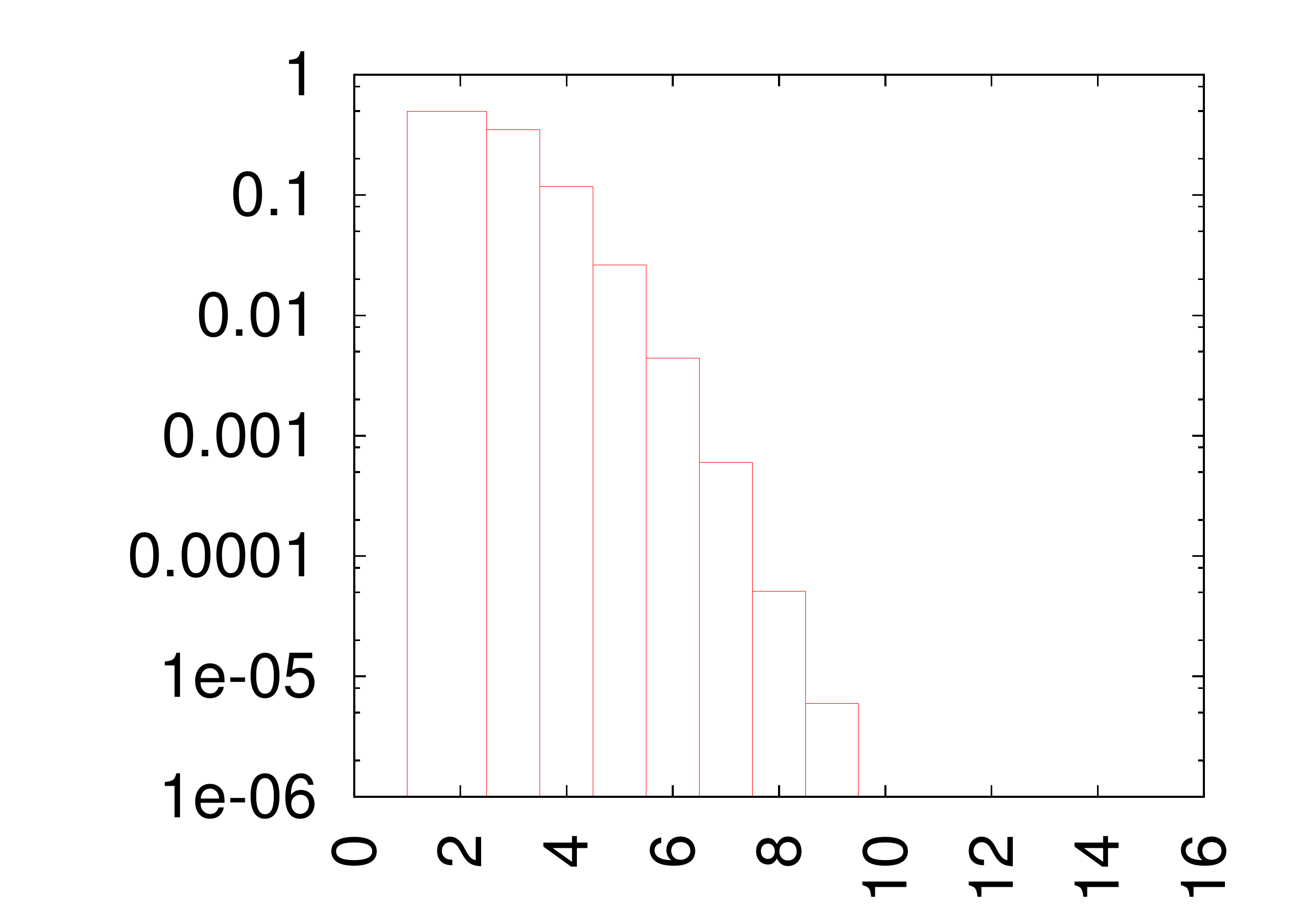}
   {\rotatebox{-90}{\kern0.2cm PATENTS-RANDOM}}
}
\caption{DWAG[(a),(b),(c)], DWAG-RANDOM[(d),(e),(f)], PATENTS[(g),(h),(i)], PATENTS-RANDOM[(j),(k),(l)]
\label{fig:plot_set1}
}
\end{figure*}

\begin{figure*}
\centering
\captionsetup[subfigure]{labelformat=empty}
\subfloat[]{%
      \small
      \begin{tabularx}{\textwidth}{p{1cm}XXX}
        & Indegree Distribution & Outdegree Distribution & RRL path length distribution
      \end{tabularx}
    }

 \subfloat[]{
   \includegraphics[width=1.40in,angle=-90]{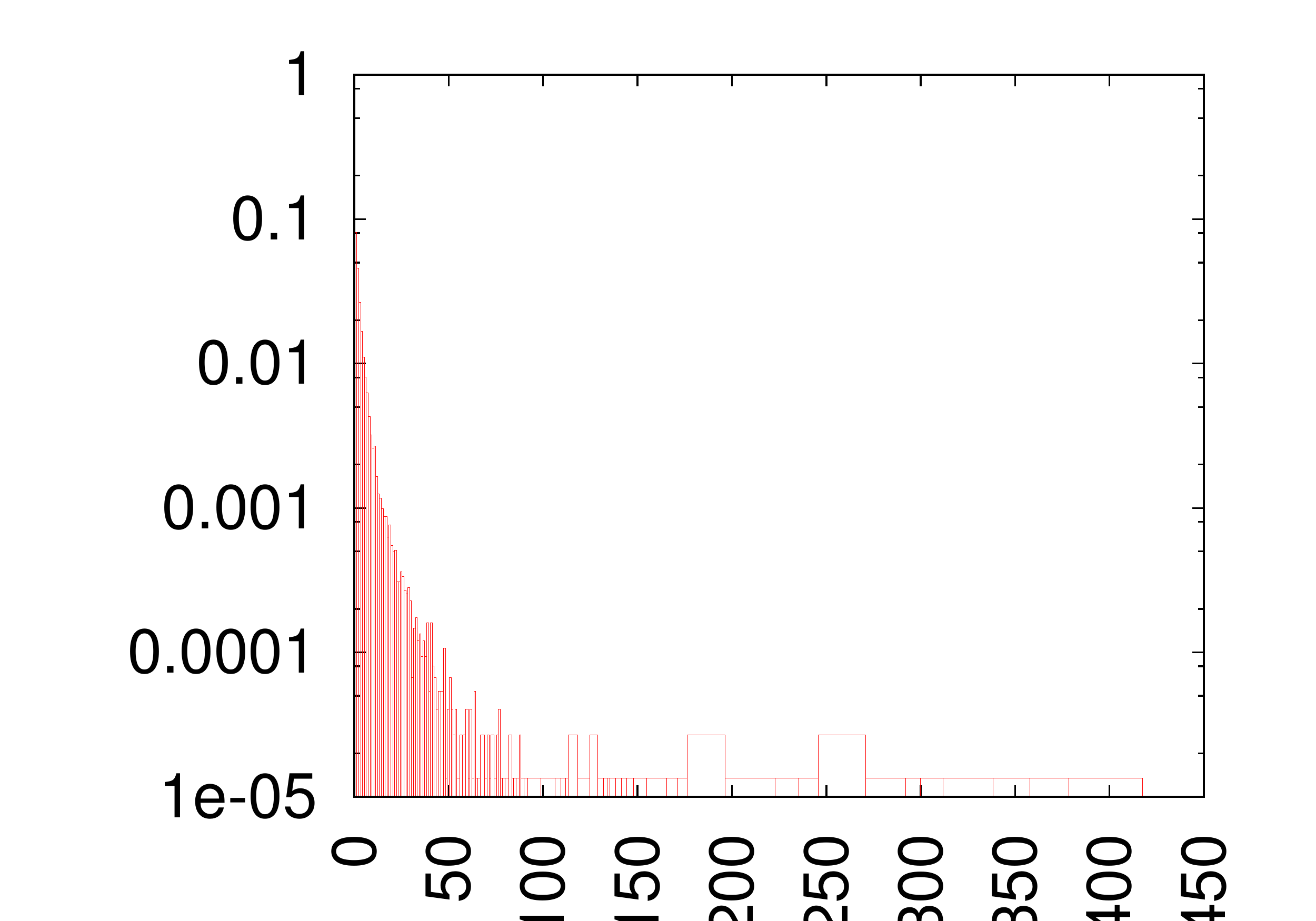}
}\hspace{-1em}
 \subfloat[]{
   \includegraphics[width=1.40in,angle=-90]{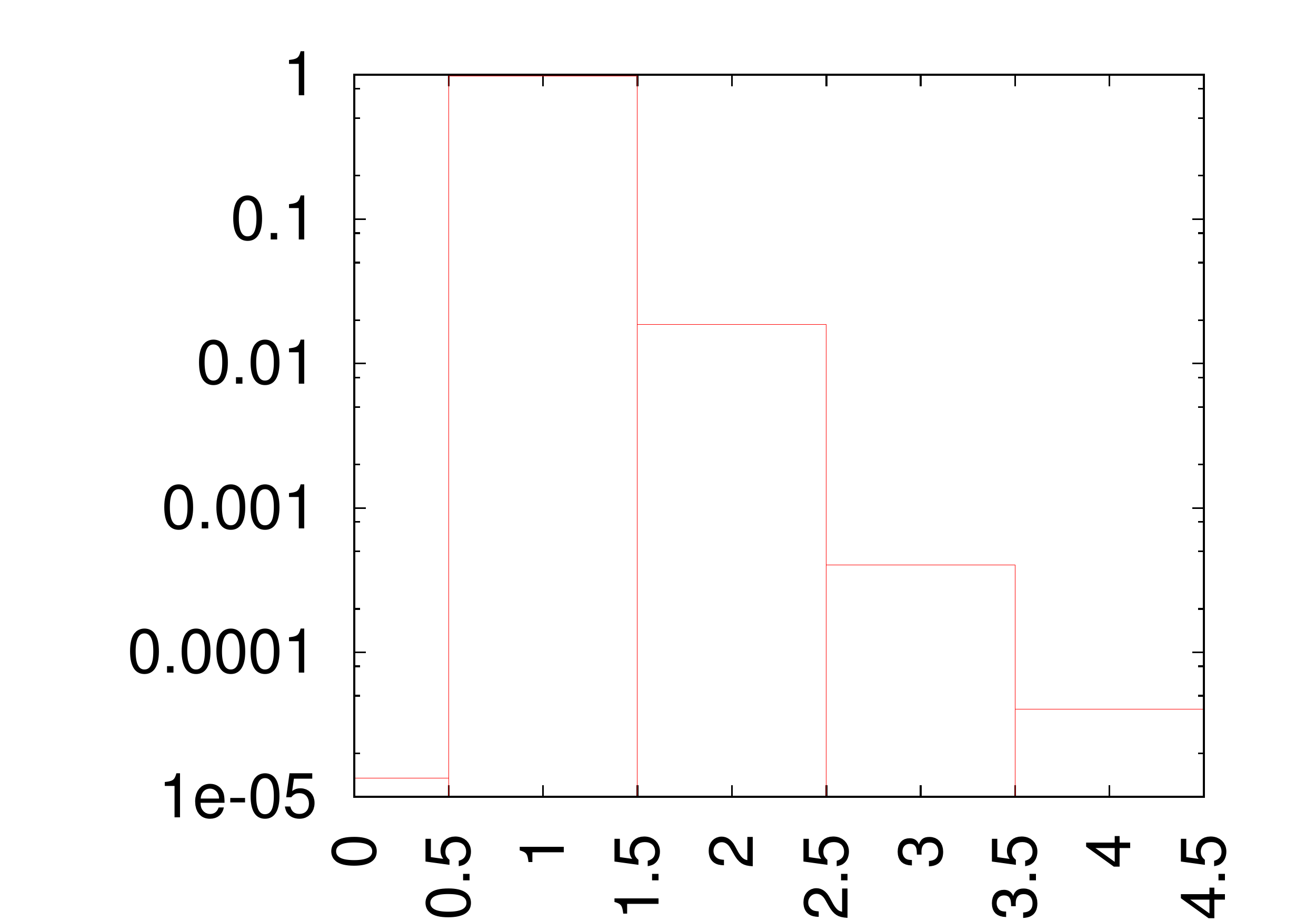}
}\hspace{-1em}
 \subfloat[]{
   \includegraphics[width=1.40in,angle=-90]{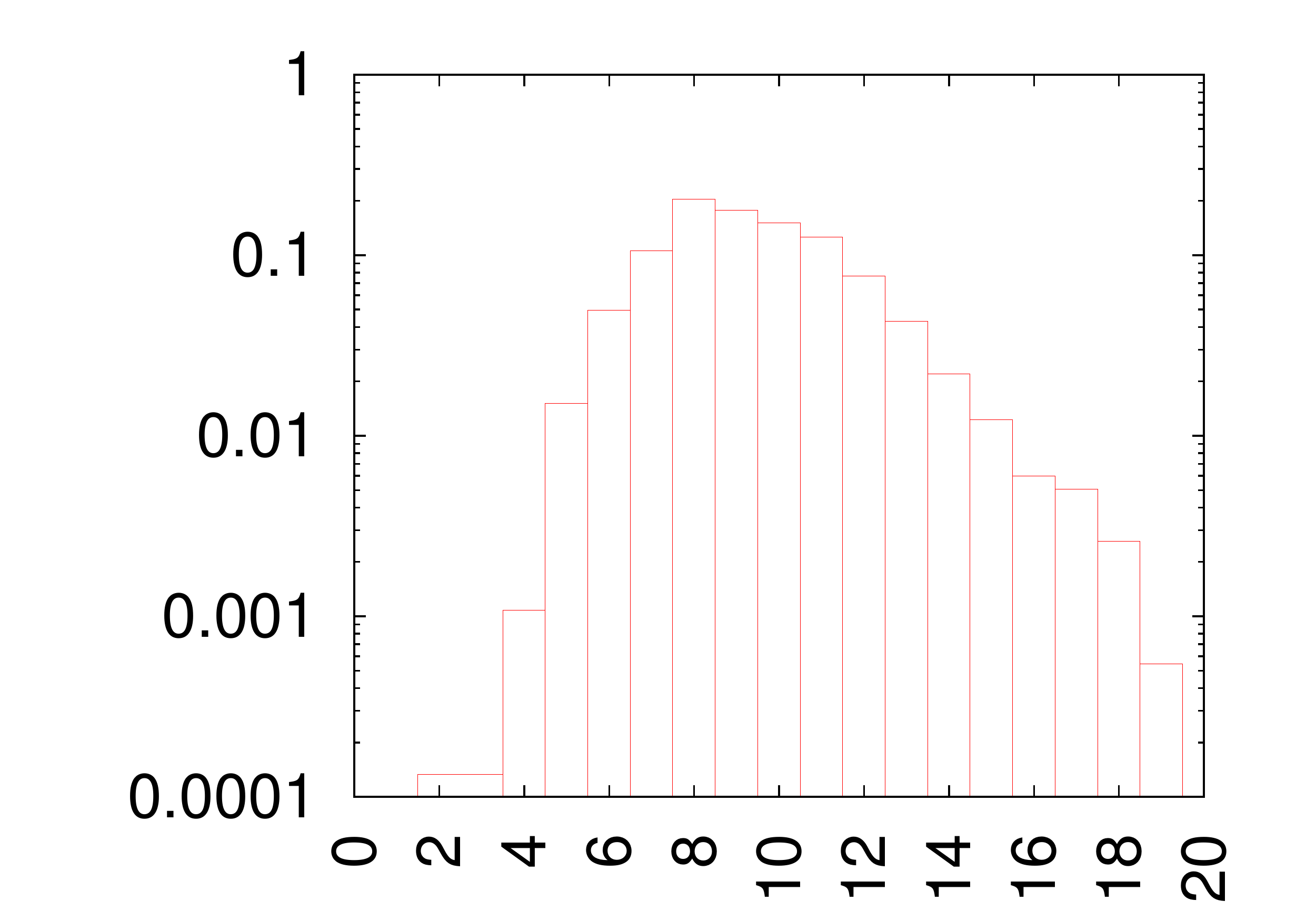}
   {\rotatebox{-90}{\kern0.6cm WORDNET}}
}
\\[-0.6cm]
 \subfloat[]{
   \includegraphics[width=1.40in,angle=-90]{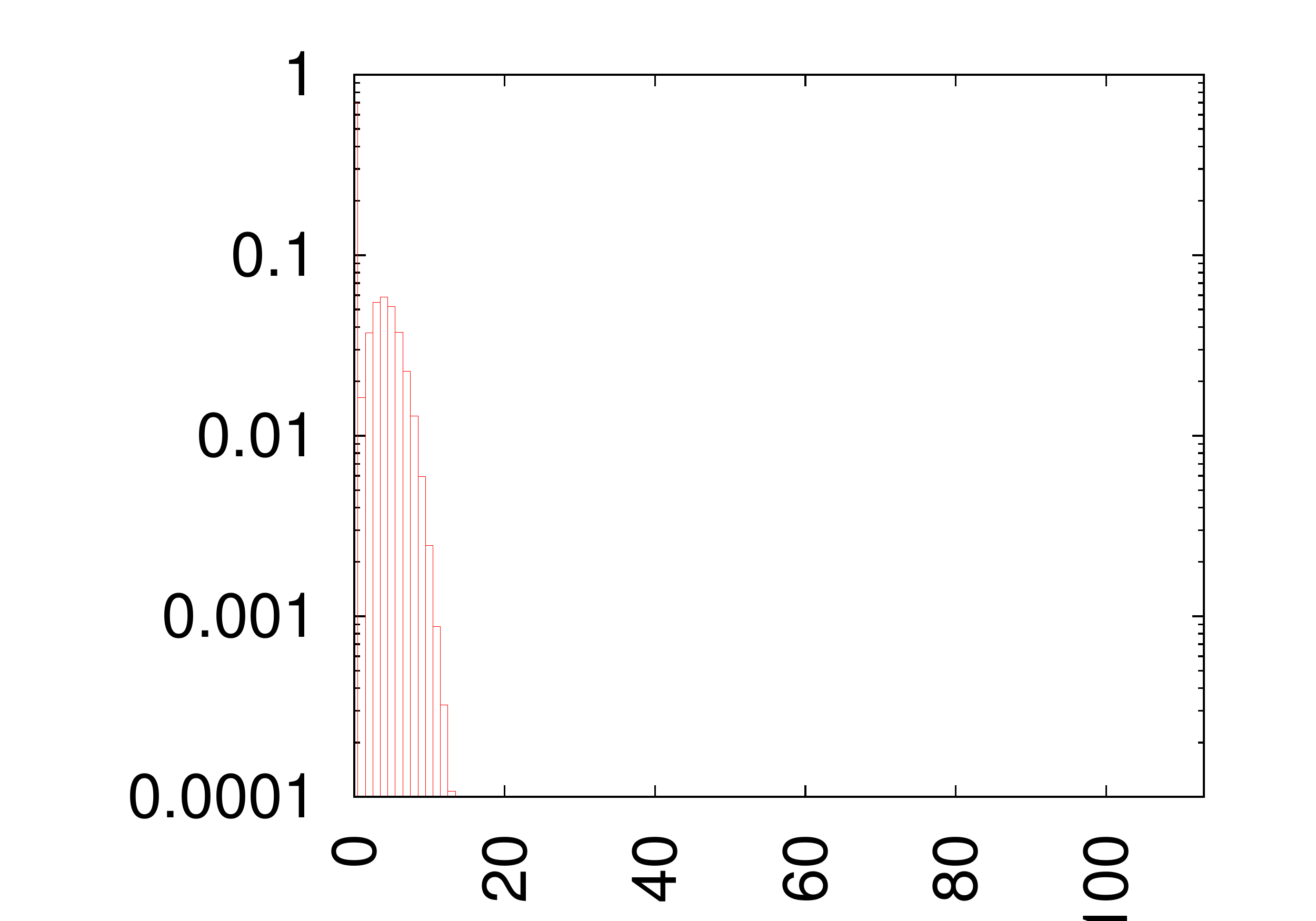}
}\hspace{-1em}
 \subfloat[]{
   \includegraphics[width=1.40in,angle=-90]{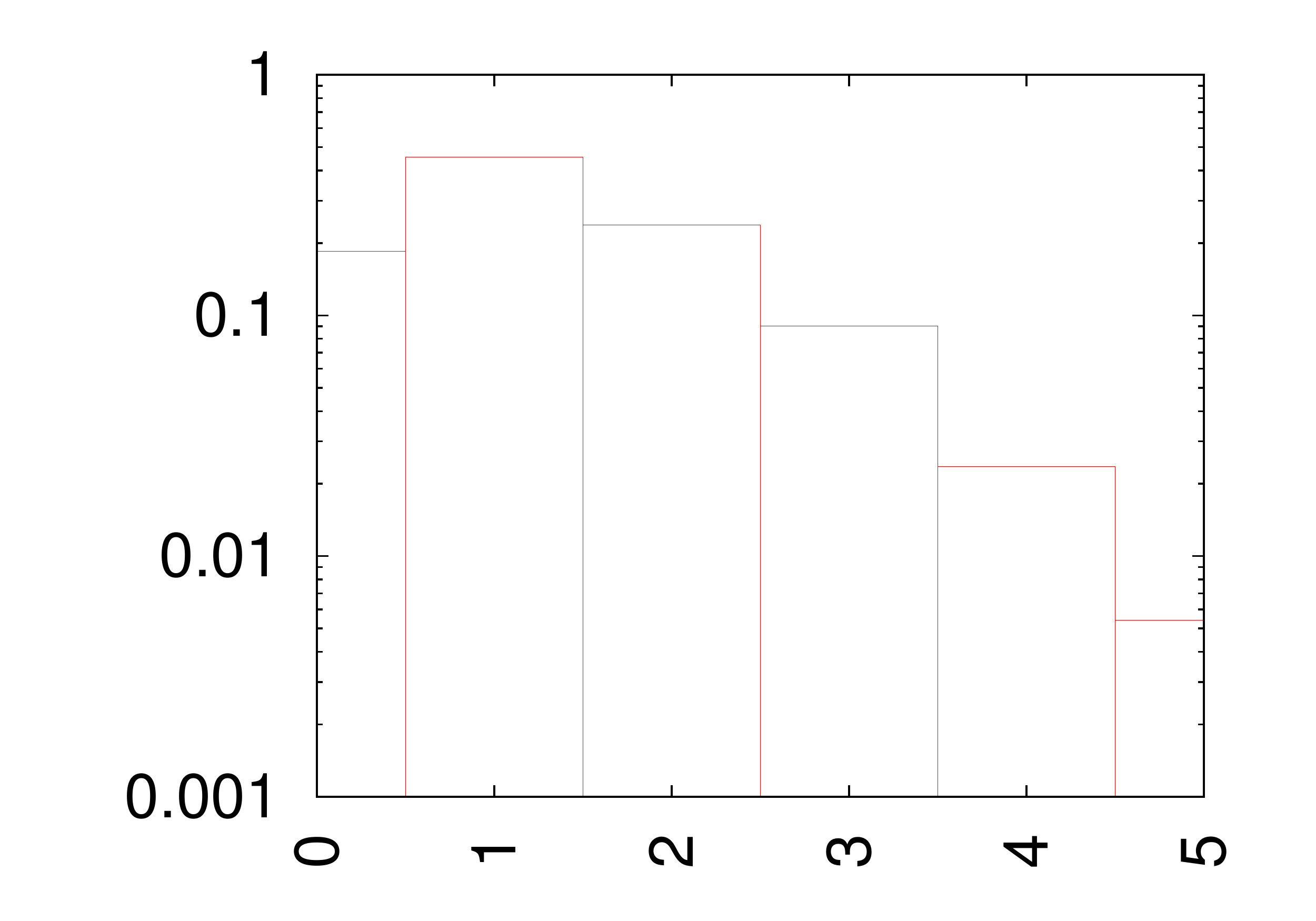}
}\hspace{-1em}
 \subfloat[]{
   \includegraphics[width=1.40in,angle=-90]{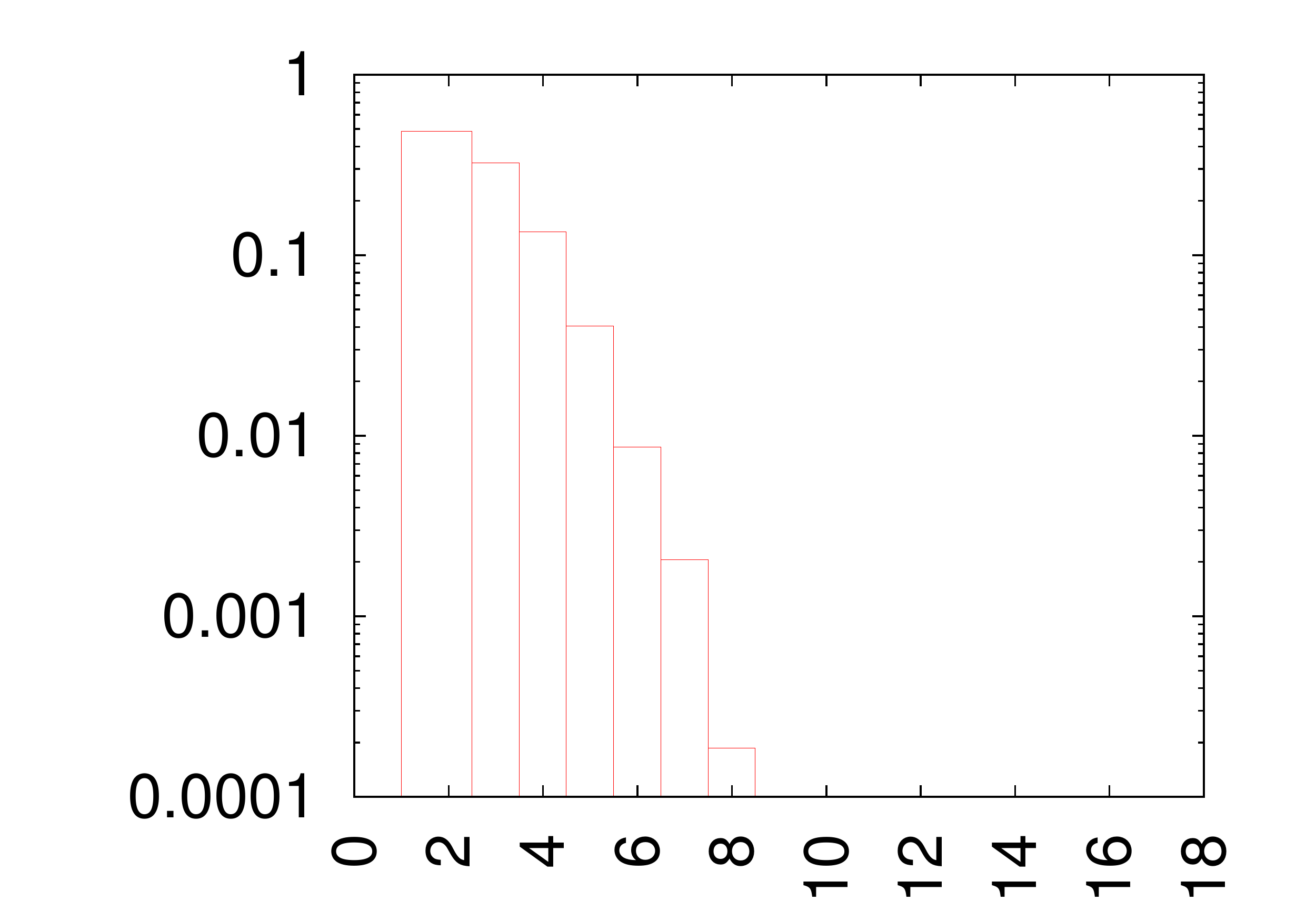}
   {\rotatebox{-90}{\kern0.1cm WORDNET-RANDOM}}
}
\\[-0.6cm]
 \subfloat[]{
   \includegraphics[width=1.40in,angle=-90]{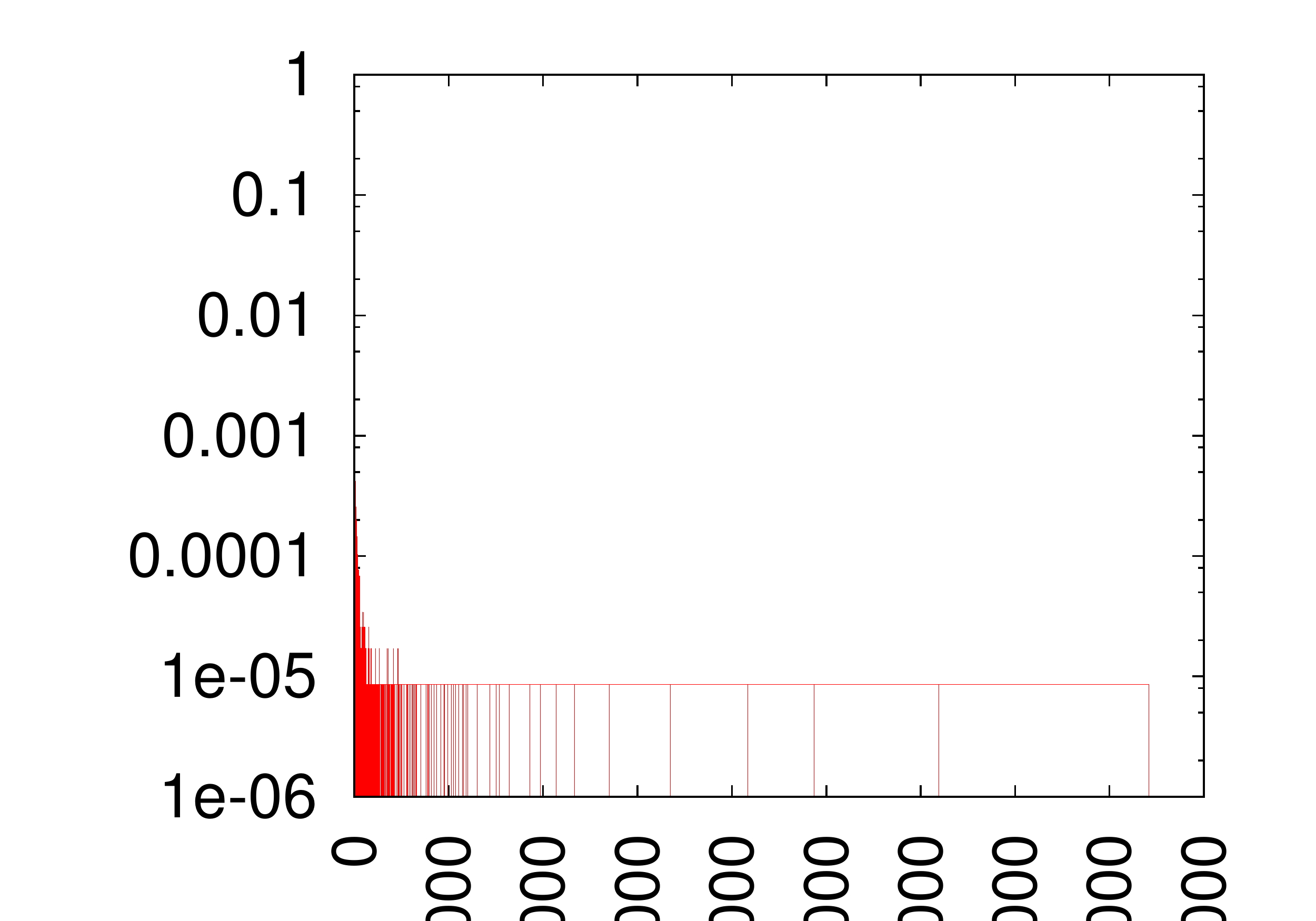}
}\hspace{-1em}
 \subfloat[]{
   \includegraphics[width=1.40in,angle=-90]{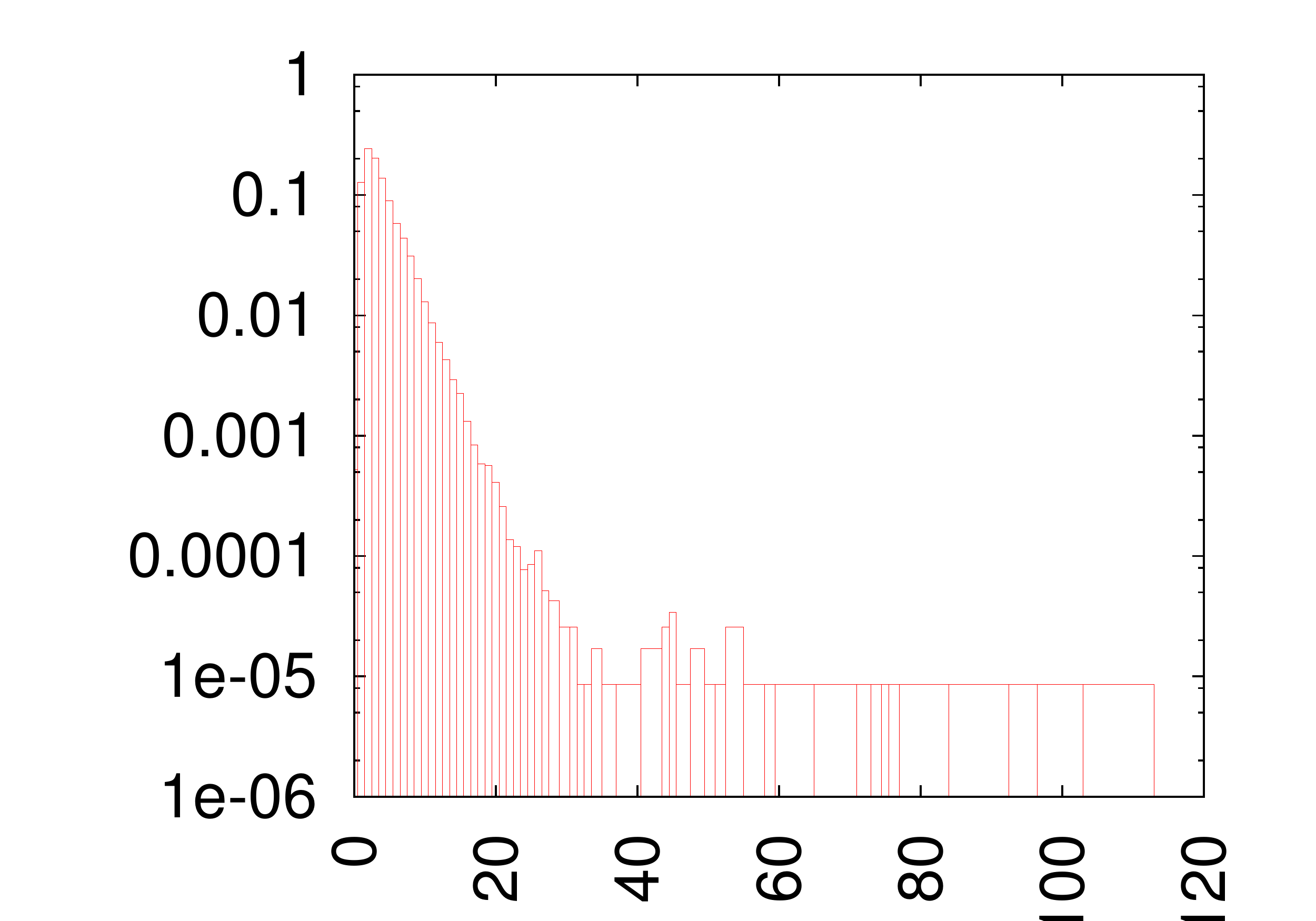}
}\hspace{-1em}
 \subfloat[]{
   \includegraphics[width=1.40in,angle=-90]{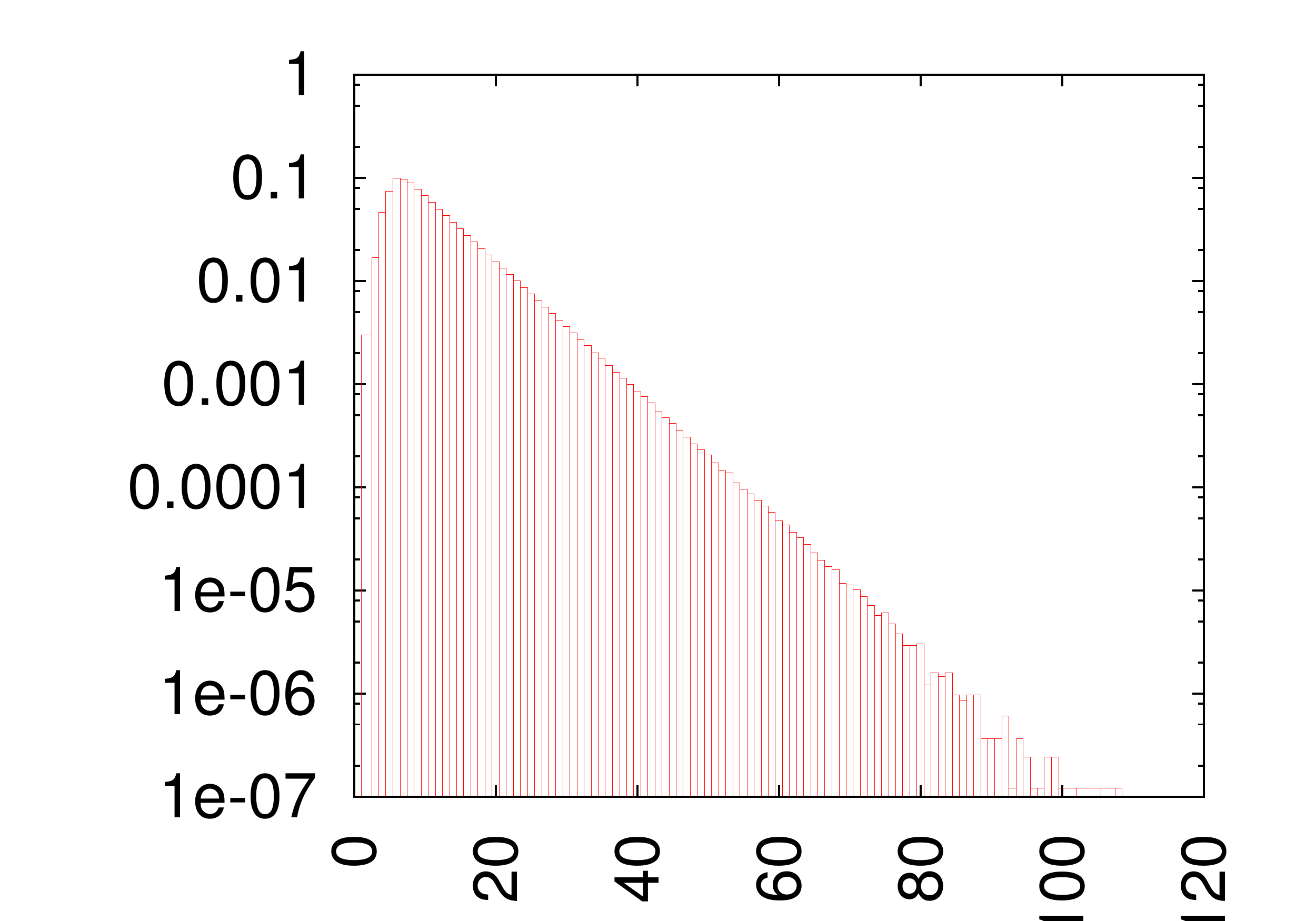}
   {\rotatebox{-90}{\kern0.5cm CYC}}
}
\\[-0.6cm]

 \subfloat[]{
   \includegraphics[width=1.40in,angle=-90]{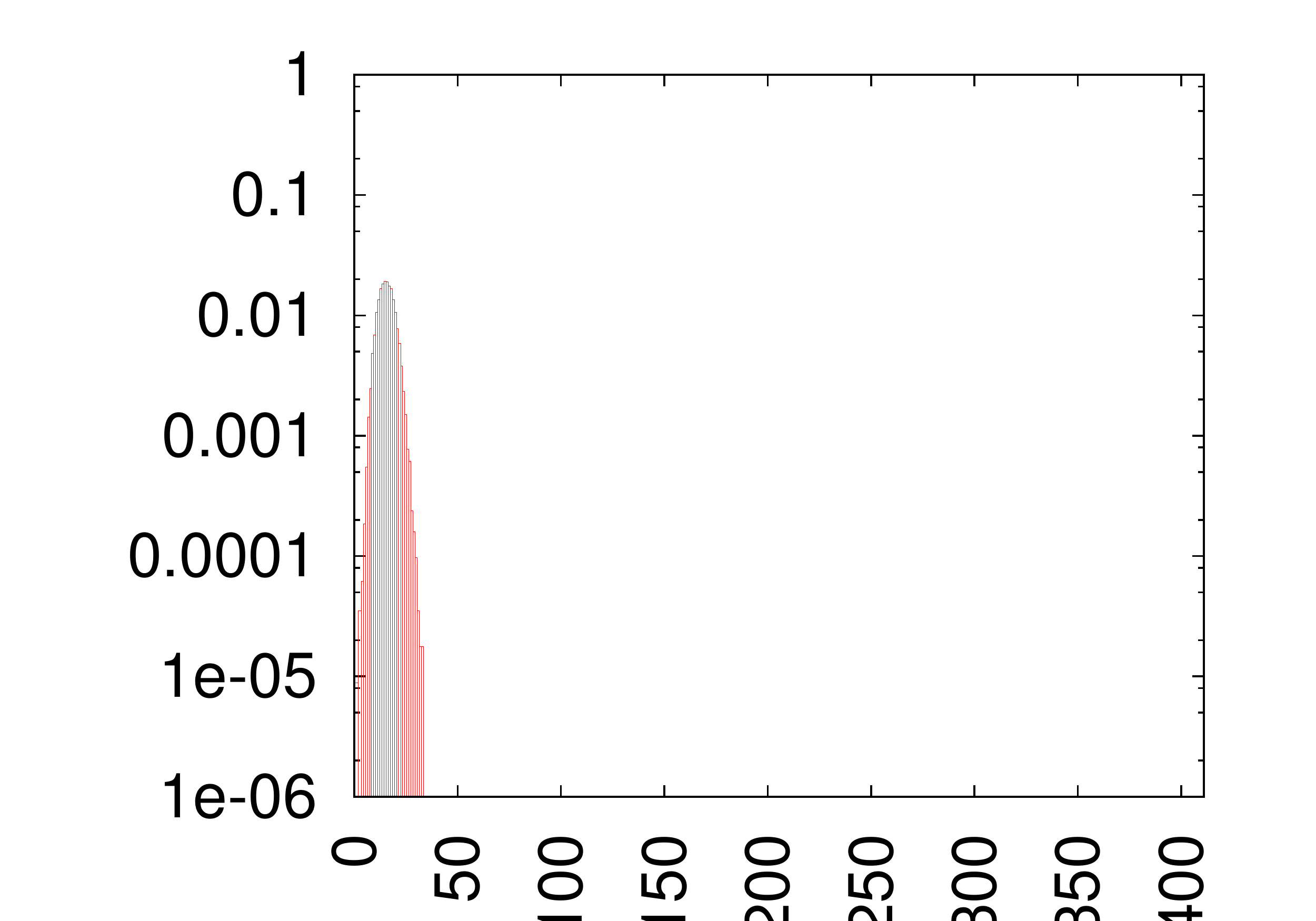}
}\hspace{-1em}
 \subfloat[]{
   \includegraphics[width=1.40in,angle=-90]{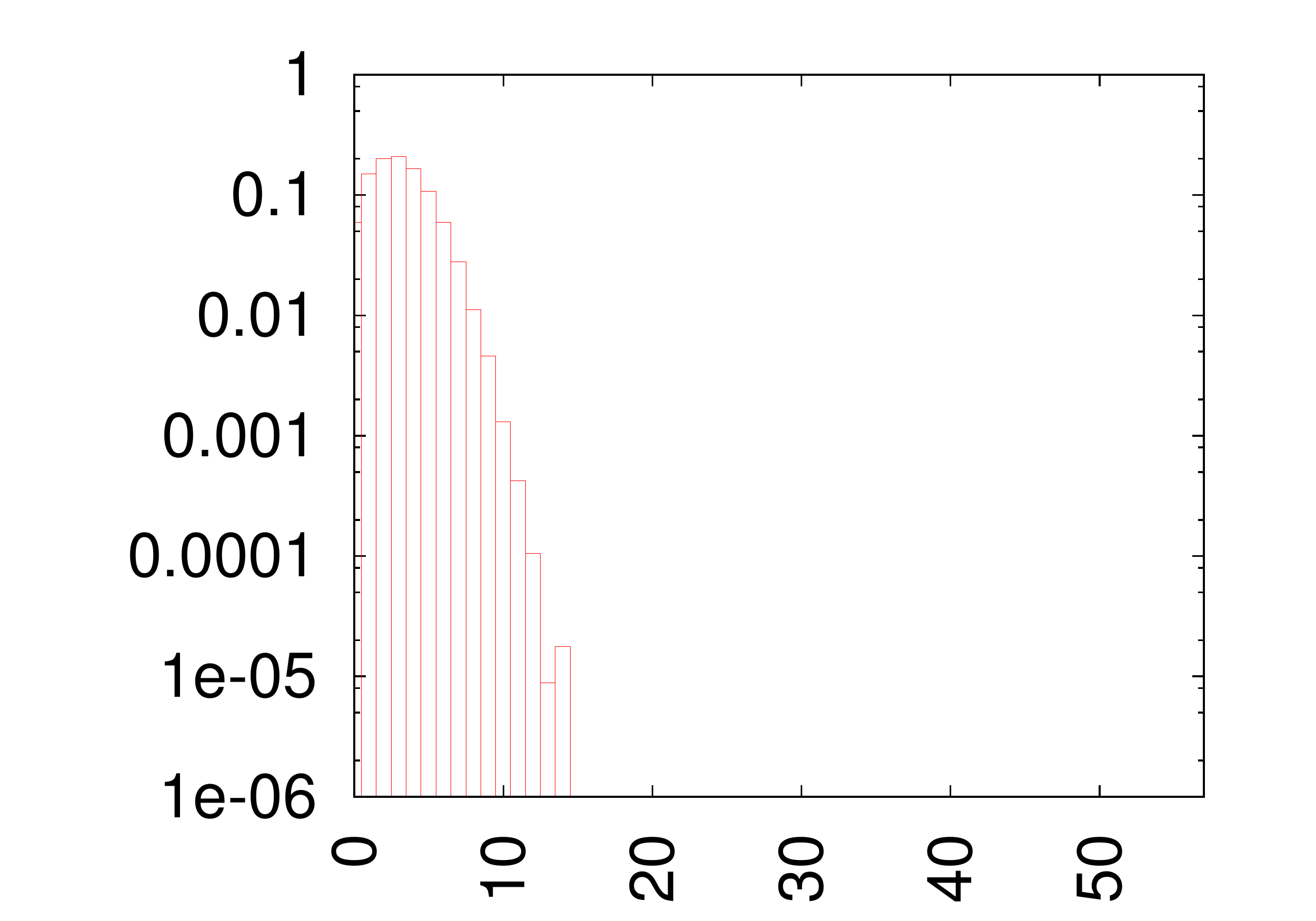}
}\hspace{-1em}
 \subfloat[]{
   \includegraphics[width=1.40in,angle=-90]{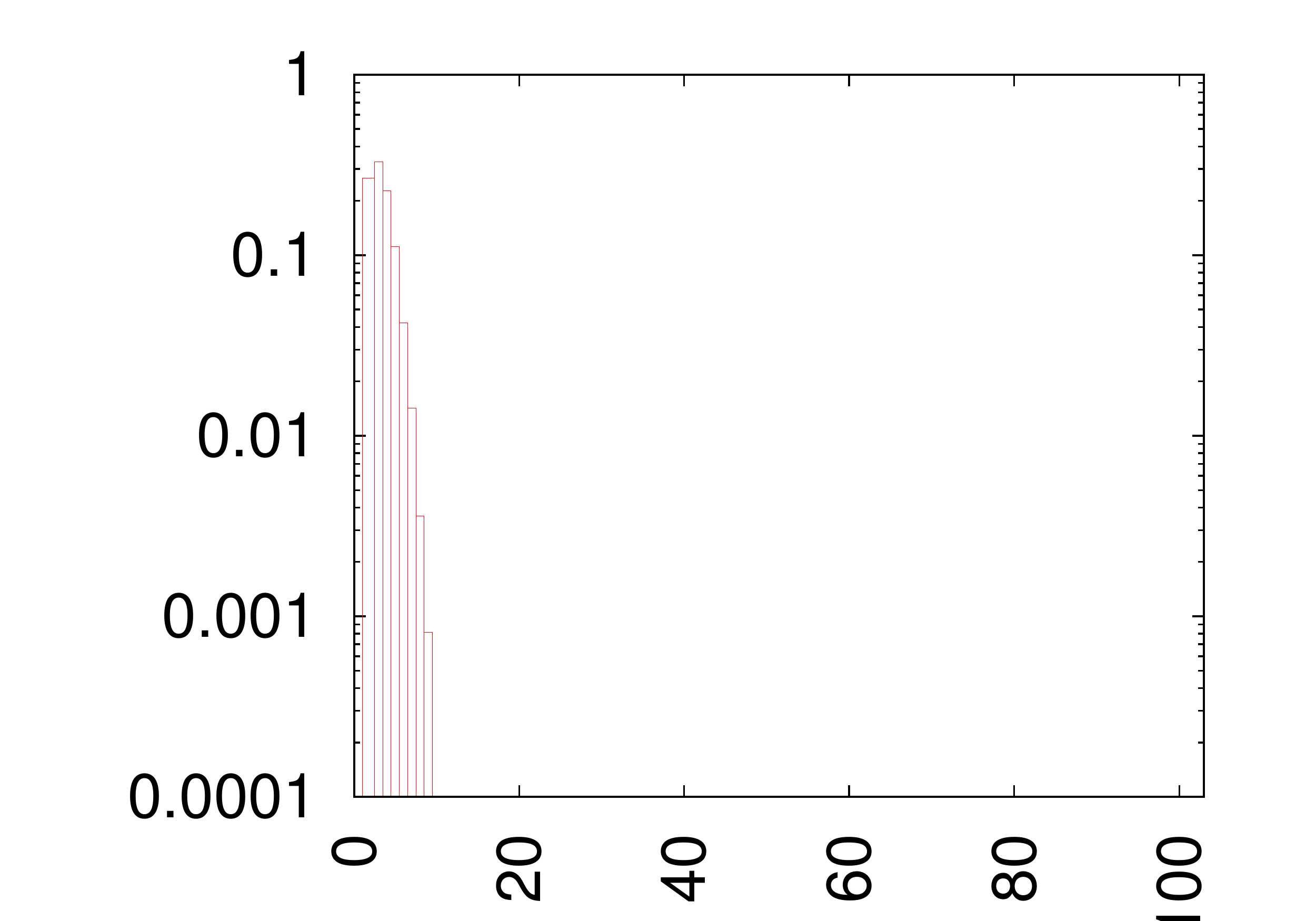}
   {\rotatebox{-90}{\kern0.1cm CYC-RANDOM}}
}
\\[-0.6cm]

 \subfloat[]{
   \includegraphics[width=1.40in,angle=-90]{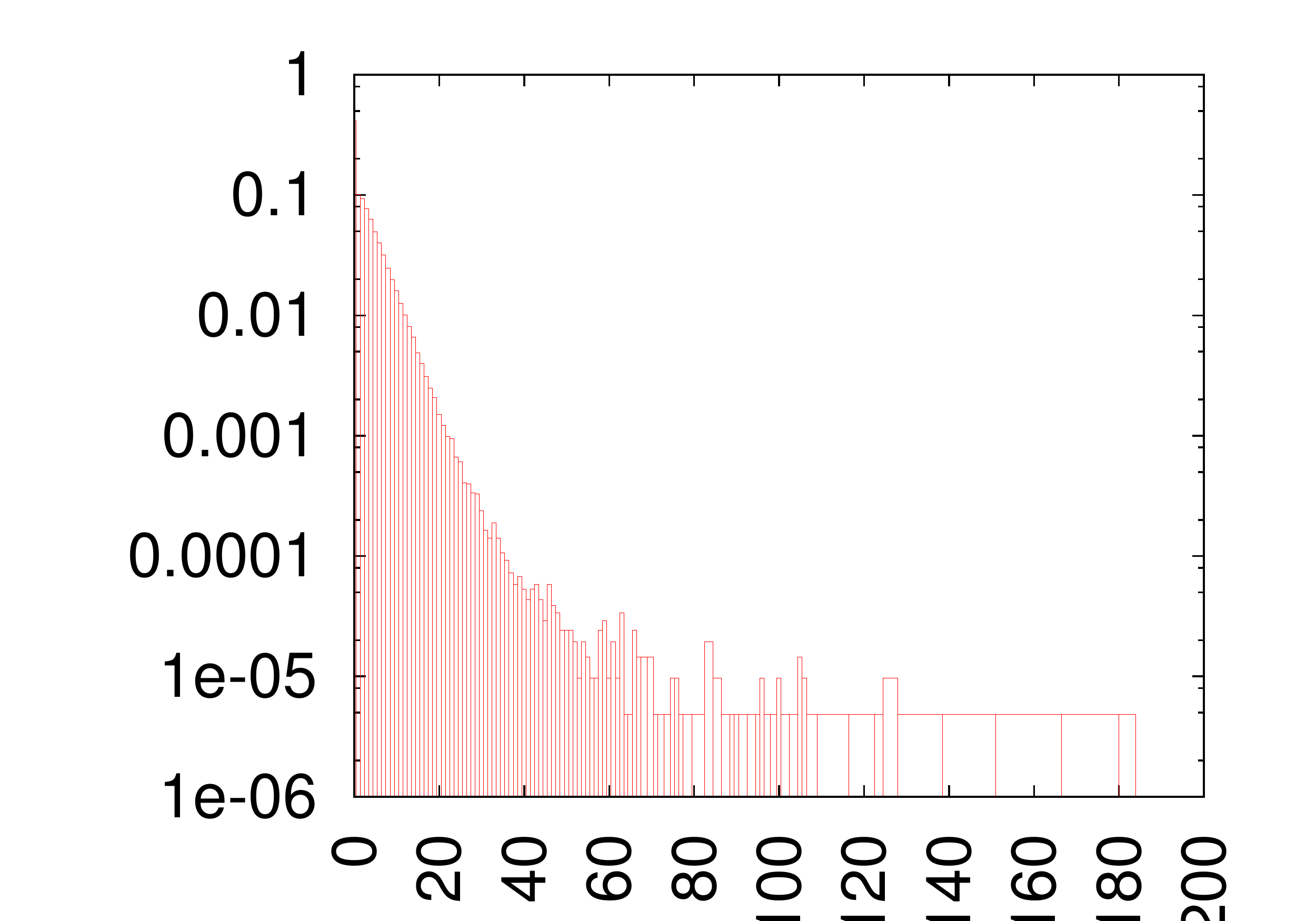}
}\hspace{-1em}
 \subfloat[]{
   \includegraphics[width=1.40in,angle=-90]{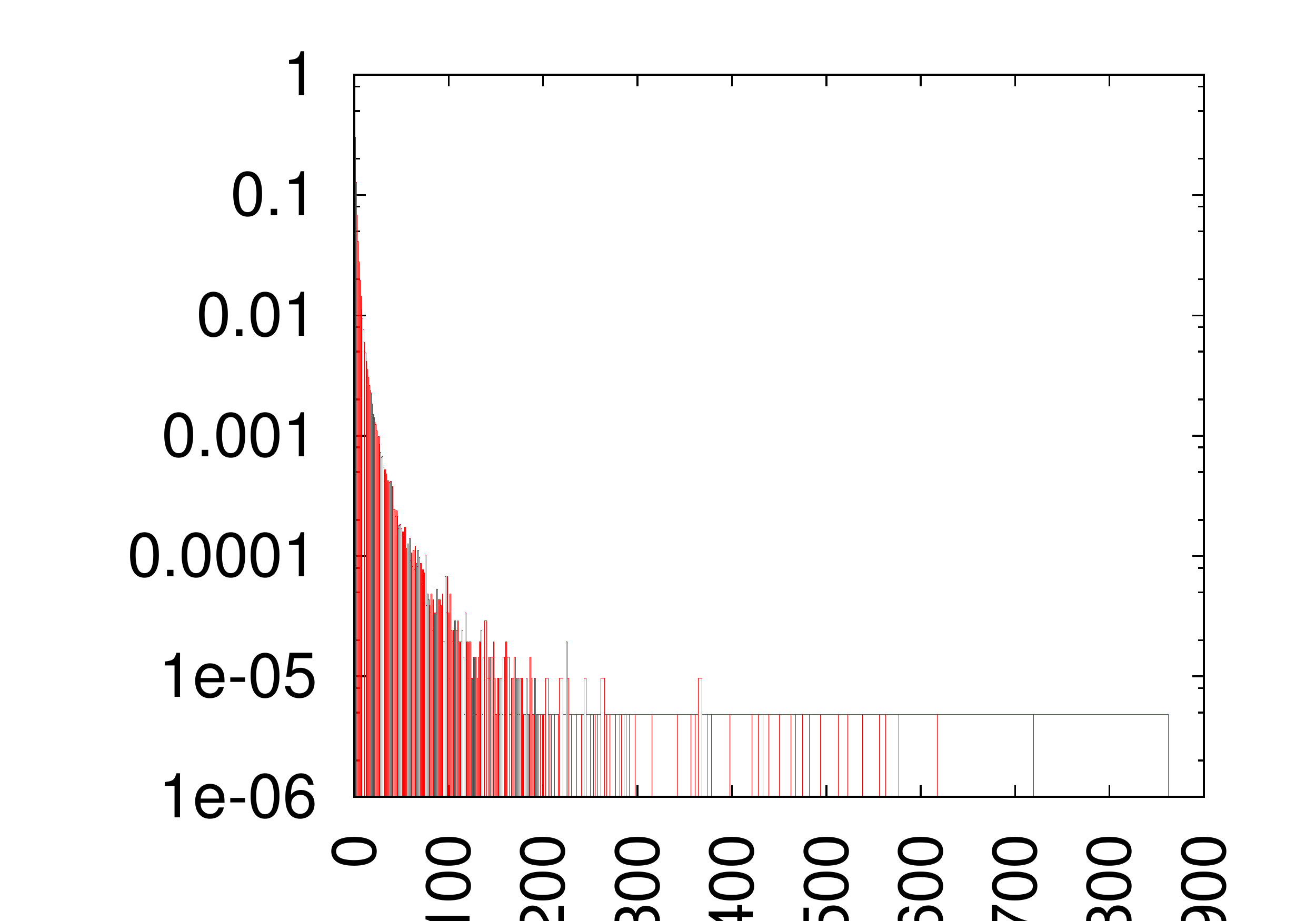}
}\hspace{-1em}
 \subfloat[]{
   \includegraphics[width=1.40in,angle=-90]{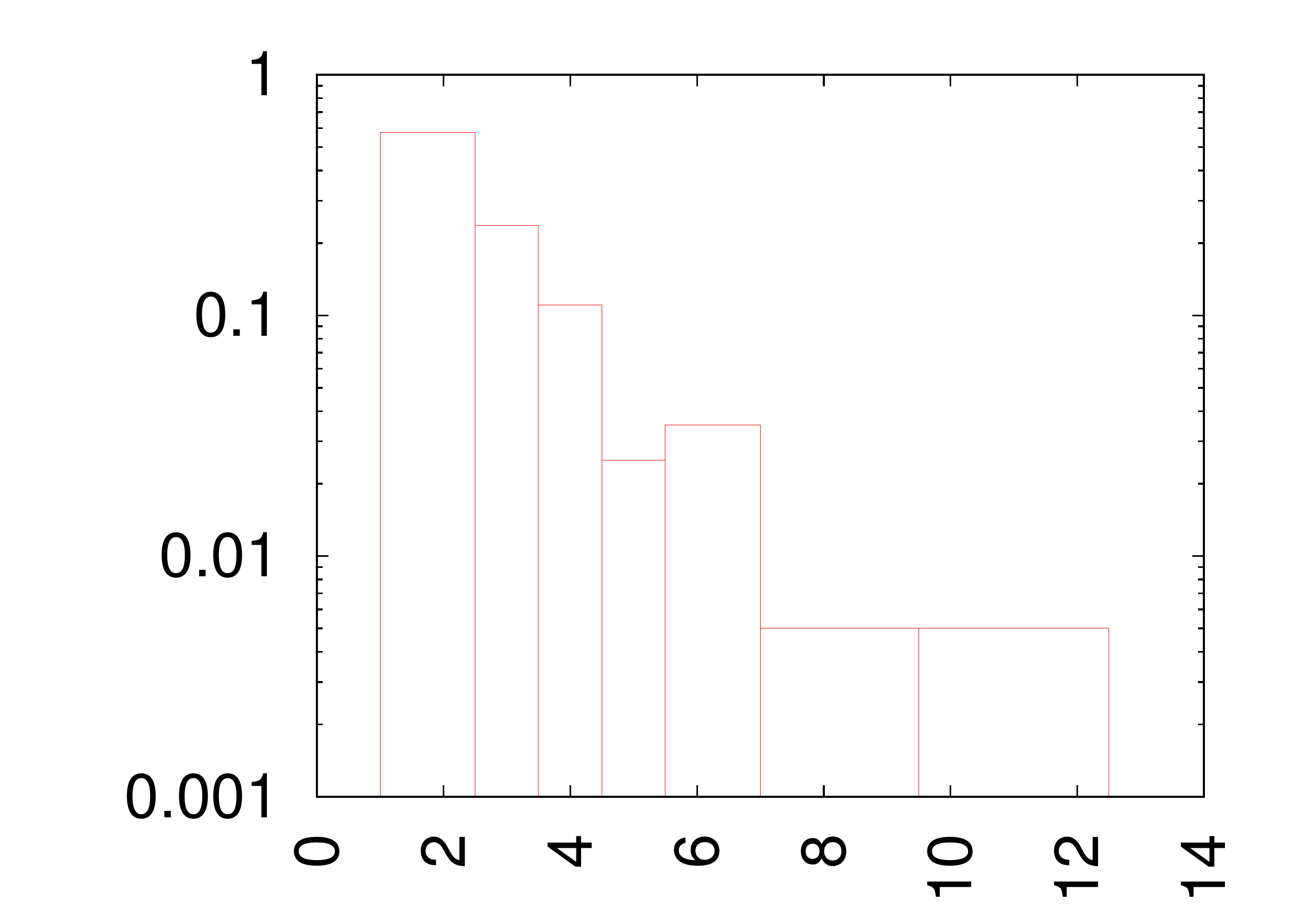}
   {\rotatebox{-90}{\kern0.1cm CITATION}}
}
\\[-0.6cm]

 \subfloat[Indeg distribution]{
   \includegraphics[width=1.40in,angle=-90]{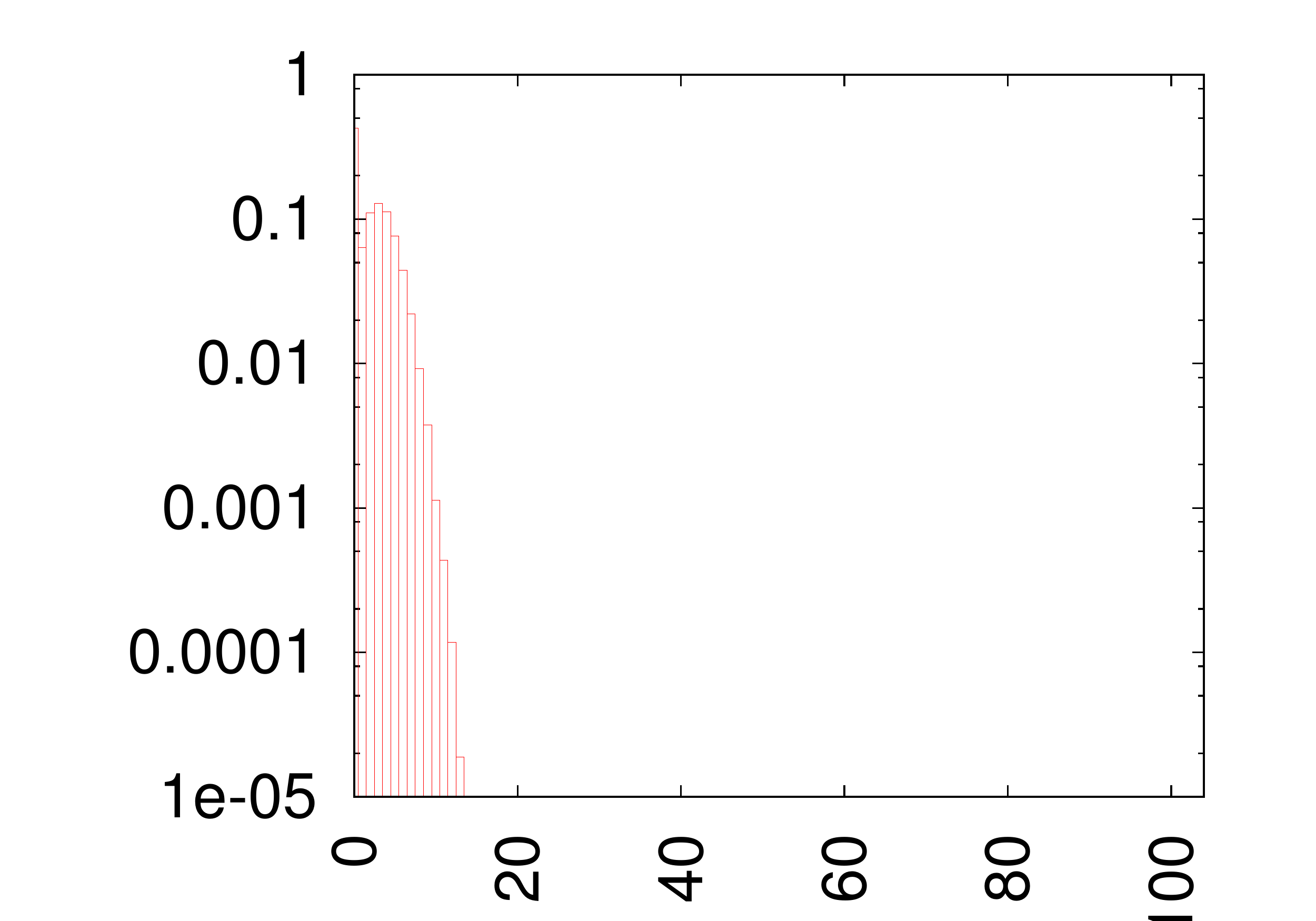}
}\hspace{-1em}
 \subfloat[Outdeg distribution]{
   \includegraphics[width=1.40in,angle=-90]{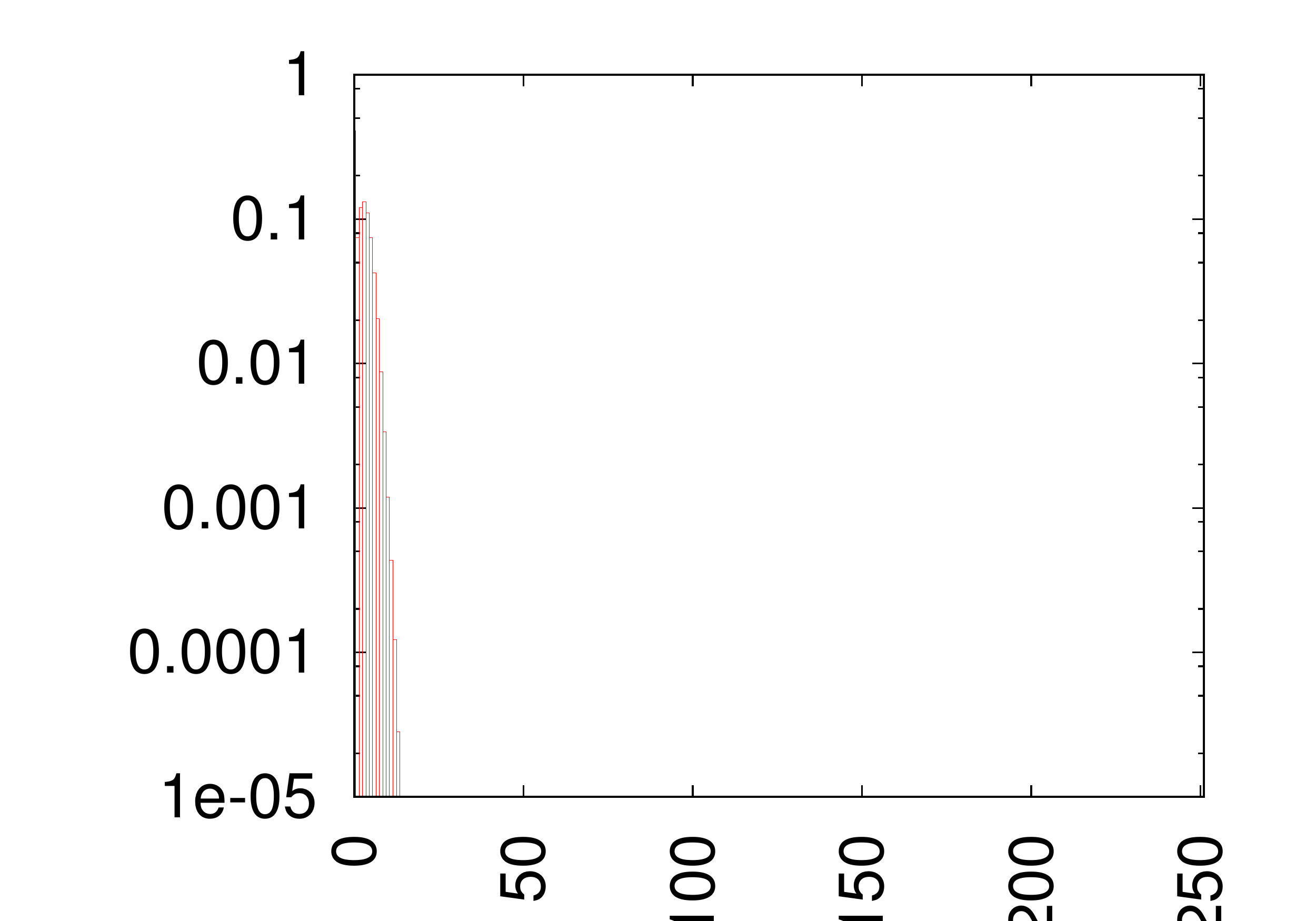}
}\hspace{-1em}
 \subfloat[Random path lenght distribution]{
   \includegraphics[width=1.40in,angle=-90]{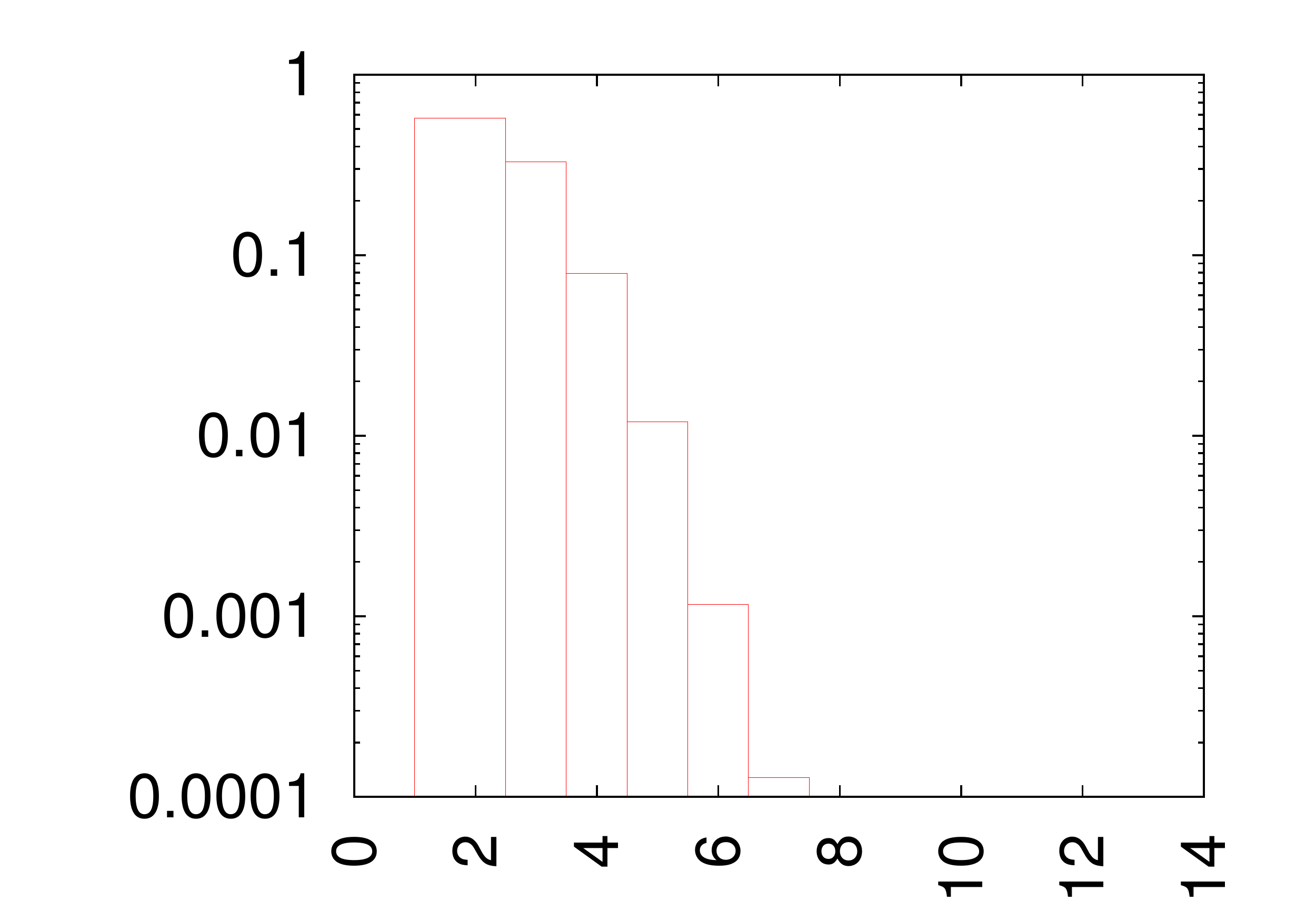}
   {\rotatebox{-90}{\kern0.1cm CITATION-RANDOM}}
}
\caption{Indegree, outdegree, and, RRL path length distributions: Real world vs. random graphs}
\label{fig:plot_set2}
\end{figure*}

\section{Comparison of Real with Random Graphs}
\label{sec:randomdag}
\begin{wrapfigure}{r}{0.05\textwidth}
\resizebox{1.5cm}{1.5cm}{
\begin{tikzpicture}[font=\Large]
 \SetUpEdge[lw         = 1.5pt,
            color      = orange,
            labelcolor = white]
  \GraphInit[vstyle=Normal] 
  \SetGraphUnit{3}
  \tikzset{VertexStyle/.append  style={fill}}
  \Vertex[x=0, y=0, L=b]{B}
  \Vertex[x=-1, y=1, L=r]{R}
  \Vertex[x=0, y=1, L=e]{E1}
  \Vertex[x=1, y=1, L=e]{E2}
  \Vertex[x=-1, y=2, L=e]{E3}
  \Vertex[x=1, y=2, L=a]{A}
  \Vertex[x=0, y=3, L=d]{D}
  \tikzset{EdgeStyle/.style={->}}
  \Edge (B)(R)
  \Edge(B)(E1)
  \Edge(B)(E2)
  \Edge(R)(E3)
  \Edge(E1)(A)
  \Edge(A)(D)
  \Edge(E2)(E3)
  \Edge(R)(E1)
  \Edge(E3)(D)
\end{tikzpicture}
}
\end{wrapfigure}
We compare the structural properties of the real world graph 
 with a random graph having similar make and features, i.e., the number of nodes, edges , roots, and, leafs of the random graph 
is approximately same as that of the corresponding real world graph. 
We studied following six real world acyclic graphs:

\begin{LaTeXdescription}
\item[DAWG] Directed  acyclic word graph  is a data structure that succinctly represents a collection of words.
The data structure facilitates efficient retrieval of  word (or words) matching a certain prefix.  Nodes in DWAG represent a alphabet.
Alphabets along a path from root to leaf form a word from the collection of the words. Figure on the right shows DAWG data structure over words {\it bred, bread, bead, beed,}.  

\item [Patents]  is an acyclic graph defined by citations among patents. Each node represent a patent and an edge represents citation from one node to another. 

\item [Wordnet-Hypernym] The acyclic graph is obtained from Wordnet,  which is a large lexical database of English words and concepts.  Embedded on these words, is a  network whose edges  represent conceptual-semantic and lexical relations such as  hypernym, hyponyms, meronym etc. 
Wordnet-Hypernym is a subgraph of the Wordnet graph induced by hypernym relationship. Since hypernym is a transitive relationship the Wordnet-Hypernym has acyclic structure. 

\item[Cyc] Cyc is an acyclic graph  obtained from the Cyc  ontology that represents common knowledge of everyday things. It is one of the largest and well constructed ontology. 

\item[CIT] Similar to Patent graph but from derived from publications. Each node in this acyclic graph is a paper and edges represent citation. 
  
\end{LaTeXdescription}

Table~\ref{tbl:acyc_real} and table~\ref{tbl:acyc_rand} studies  the following key characteristics for each of the above mentioned graphs:
\begin{LaTeXdescription}
\item[num nodes] : number of nodes in the graph
\item[edge factor] : average number of edge per node i.e $\frac {\text{num edges}} {\text {num nodes}}$
\item[root factor] : fraction of nodes of total num nodes having in-degree zero
\item[leaf factor] : fraction of nodes of total num nodes having out-degree zero
\item[out degree distribution's]  mean and variance  of the out degree distribution
\item[in degree distribution's] mean and variance of the in degree distribution. 
\end{LaTeXdescription}

These characteristics vary across real world graphs. Edge factor ranges
from 2.04 to 8.75. The root factor is as small as .00004 for DWAG
and as high as .81 for Cyc. 
Mean out-degree and in-degree, in comparison, is less varied  
across the graphs. They vary within the range of 1 to 4. 
The larger  mean in-degree value (and including in-degree variance) 
the more the farther away the graph structure is from a  tree structure, i.e.,
lot more edges need to be removed to make it a  tree structure. 

Compared to mean, the variance, however, has varies widely.
Variance of out-degree distribution is as low as 0.02 for Wordnet-Hypernym
to as high as 106.40 for CIT.
Similarly for in-degree,  DWAG graph has variance of 141497.
In this graph almost all nodes connect to the leaf node, whereas rest of nodes
have mostly 2 -- 3 indegree leading to such high variance.

Plots in figure~\ref{fig:plot_set1} and~\ref{fig:plot_set2} compares characteristics of these   graphs with their random graph counterpart. 
As expected,  the out-degree and in-degree of random graphs is always the same (some form of 
binomial distribution). This is in contrast to the degree distribution of the real world graphs 
whose shape vary.  Degree distribution of Wordnet-Hypernym varies between 1 to 45000 and has a sharp knee, i.e.,
almost all nodes have degree between 0 -- 5000 but few nodes have exponentially high degree.  
The out degree, however, has mild variation between 0 -70 and decays gradually. 
The in-degree and out-degree of the random graph in contrast varies in much smaller range and the decay pattern
does not follow the original graph. 

In other graphs too,  degree distribution of random graph, which has smooth decay over a small range does not conform 
to degree distribution of real world graphs which has mostly have wider range and decay exponentially. 

We now discuss structural richness  of real world graphs when compared with random graphs.
We use distribution of length of random walk from root to leave (or RRL) as a measure of its structural 
complexity. We classify simple (or degenerate graph) as those whose RRL distribution varies in a narrow range with high concentration at near zero i.e. the distribution 
drops exponentially with increase in length. 
On the other hand, RRL distribution of graphs  with irregularity and rich structure vary across a wider range and can have binomial or exponential
or any other arbitrary shape.

We see that walk length of random graph is mostly less  than 10. 
By this measure, random graphs are almost 
structured i.e. they can be seen as k-partite graph with very few edges (1/10000 fraction of total number of edges) 
accounting for irregularity.  This observation matches with the theoretical proof given earlier. 

Real world graphs show very different behavior with respect to random walk length distribution.  The shape is binomial over
a wider range. Walk length in Cyc graph can be as high as 120. 
CIT is the only graph with low walk length and therefore a random acyclic generator would be model it. 
For all other graphs random generators turn out not be good model either  for for degree distribution or their inherent structural richness.




\section{Conclusion}

Based upon the random walk length distribution, which measures the irregularity of a acyclic graph, the paper shows
that graphs produced by random generation process are structured and predictable. We also theoretically show that
random graph produce degenerate collapsed graph, i.e., with high probability the depth of dag is very small. 
We experimentally verify this for various graph generated randomly, whose other characteristic are matched with 
the real world graph. 
The paper also shows that unlike  random graphs,  real worlds graph vary in their in- and out-degree 
distribution and have much higher magnitude of irregularity as measured by random walk length distribution.

\bibliography{sig-alternate}
\bibliographystyle{abbrv}


\end{document}